\documentclass[letterpaper,onecolumn,11pt,unpublished]{quantumarticle}
\pdfoutput=1
\usepackage[utf8]{inputenc}
\usepackage[english]{babel}
\usepackage[T1]{fontenc}
\usepackage{amsmath}
\usepackage{hyperref}

\usepackage[numbers,sort&compress]{natbib}

\usepackage{tikz}

\usepackage{lipsum}

\usepackage{mathtools} 
\usepackage{comment}
\usepackage{float}
\usepackage{cancel}

\usepackage{csquotes}

\usepackage{graphicx,caption}
\usepackage[framemethod=tikz]{mdframed} 

\usepackage{tikz}
\usetikzlibrary{calc}
\usetikzlibrary{arrows}
\usepackage{xspace}
\usetikzlibrary{decorations.pathmorphing}
\usetikzlibrary{decorations.pathreplacing} 

\input{diagram.sty}

\begin{document}

\title{A diagrammatic language for the Causaloid framework}

\author{Nitica Sakharwade}
\affiliation{Perimeter Institute of Theoretical Physics, Waterloo, Canada}
\affiliation{Department of Physics and Astronomy, University of Waterloo, Waterloo, Canada}
\affiliation{Department of Physics, University of Naples Federico II, Naples, Italy}
\orcid{0000-0002-8208-9672}
\author{Lucien Hardy}
\affiliation{Perimeter Institute of Theoretical Physics, Waterloo, Canada}

\maketitle

\begin{abstract}
The Causaloid framework \cite{hardy2005probability, hardy2007towards,hardy2009qgcomputers} is an operational approach aimed to house both the radical aspects of General Relativity -- dynamic causal structure, and Quantum Theory -- indefiniteness, to provide a scaffolding that might be suitable for Quantum Gravity by providing a landscape of theories that allow for \emph{indefinite causal structure}. One may consider it as a generalisation of generalised probability theories (or GPTs) where {\it a priori} regions are not assumed to have any given causal relationship, to incorporate the possibility of indefinite causal structure.  Since its conception, there have been many advances in the field of indefinite causal structure mostly stemming from the work of Chiribella et al.\ on the quantum switch \cite{chiribella_quantum_2013} and supermaps and from Oreshkov et al.\ on causal inequalities and process matrices \cite{oreshkov2012quantum}.  These approaches have systems moving along wires and use a Hilbert space structure.  They violate the standard causality constraints of Quantum Theory and, in this sense, can be regarded as \emph{post-quantum}. The Causaloid approach does not necessarily have systems moving along wires or Hilbert spaces.  
This is the first paper in a trilogy of papers aiming to close the gap between the Causaloid (that allows for GPTs) and post-quantum studies that employ Hilbert spaces. To do so in the present paper, we provide a diagrammatic language for the Causaloid framework along with new terminology for the three levels of physical compression (called Tomographic, Compositional, and Meta compression). 
\end{abstract}

\section{Introduction}
The Causaloid framework \cite{hardy2007towards,hardy2005probability,hardy2009qgcomputers} suggests a research program aimed at finding a theory of Quantum Gravity. On one side General Relativity (GR) while deterministic, features dynamic causal structures; on the other side Quantum Theory (QT) while having fixed causal structures, is inherently probabilistic in nature. This inherent probabilistic nature is associated with indefiniteness wherein there is no matter-of-the-fact as to the value of some physical variables. It is natural to then expect Quantum Gravity (QG) to house both of the radical aspects of GR and QT and therefore incorporate \emph{indefinite causal structure} (this term first appeared in \cite{hardy2009qgcomputers}). Further, the Causaloid framework is based on a very basic operational methodology -- the assertion that {\it any physical theory, whatever it does, must correlate recorded data}. One may think of enforcing an operational methodology through a narrative of imagining a person inside a closed space, having access to stacks of cards with recorded data (procedures, outcomes, locations); and the person is tasked with inferring (aspects of) the underlying physical theory that governs the data. The correlation within recorded data due to the physical theory implies that the stacks of cards are filled with (some) redundancy. The person in the box distils away the redundancy by \emph{compressing} the data. This is called \emph{Physical Compression} since it is governed by the nature of the underlying physical theory. In this framework there are three levels of compression: 1) Tomographic Compression - the compression that allows us to write down the state by performing only a few procedures and finding their outcomes 2) Compositional Compression - the compression that allows the composition of causally adjacent processes to be described as a new compressed process and 3) Meta Compression - the compression that captures the structure of the tomographic and compositional compression over different regions.

\begin{figure}[h!]
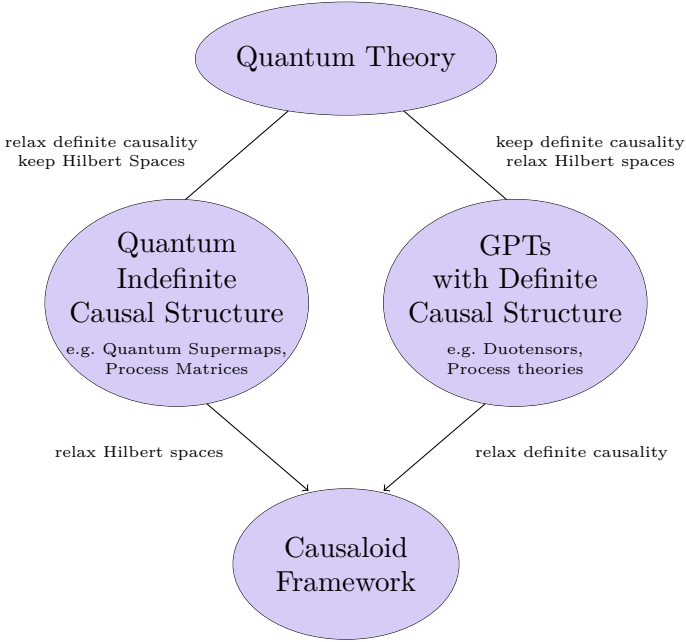

\begin{center}
\begin{Diagram}{3}{-20}
\begin{move}{0,0}
\draw [->] (-1,11) -- (-9,4.1);
\draw [->] (1,11) -- (9,4.1);
\draw (0,12) ellipse (8 and 3);
\fill[blue!80!red!20!] (0,12) ellipse (8 and 3);
\draw (0,12) node {Quantum Theory};
\draw (-13,6.5) node {\tiny keep Hilbert Spaces};
\draw (-13,7.5) node {\tiny relax definite causality};

\draw (13,6.5) node {\tiny relax Hilbert spaces};
\draw (13,7.5) node {\tiny keep definite causality};
\draw [->] (-9,-5) -- (-2,-11);
\draw [->] (9,-5) -- (2,-11);
\draw (12,-9) node {\tiny relax definite causality};
\draw (-12+1,-9) node {\tiny relax Hilbert spaces};
\draw (-9,-1) ellipse (7 and 5.5);
\fill[blue!80!red!20!] (-9,-1) ellipse (7 and 5.5);
\draw (-9,2.1) node {Quantum};
\draw (-9,0.3) node {Indefinite};
\draw (-9,-1.5) node {Causal Structure};
\draw (-9,-3.5) node {\tiny e.g.\ Quantum Supermaps,};
\draw (-9,-4.5) node {\tiny Process Matrices};
\draw (9,-1) ellipse (7 and 5.5);
\fill[blue!80!red!20!] (9,-1) ellipse (7 and 5.5);
\draw  (9,2.1) node {GPTs};
\draw  (9,0.3) node {with Definite};
\draw  (9,-1.5) node {Causal Structure};
\draw (9,-3.5) node {\tiny e.g.\ Duotensors,};
\draw (9,-4.5) node {\tiny Process theories};
\draw (0,-14.9) ellipse (6 and 4);
\fill[blue!80!red!20!] (0,-14.9) ellipse (6 and 4);
\draw (0,-14) node {Causaloid};
\draw (0,-15.8) node {Framework};
\end{move}
\end{Diagram}
\caption{How general is the Causaloid framework? In this figure we see that the Causaloid framework can be thought of as a Generalised Probability Theory (GPT) for indefinite causal structure that is capable of accommodating quantum theories having indefinite causal structure (such as Quantum Supermaps \cite{chiribella_quantum_2013} and process matrices \cite{oreshkov2012quantum}) on the one hand and GPTs with definite causal structure \cite{hardy2001quantum,barrett2007information,chiribella2010probabilistic} on the other.\label{fig:GPTplusICS} }
\end{center}
\end{figure}

Since the conception of the Causaloid framework, there have been studies focusing on indefinite causal structures from the perspective of post-quantum theories based on finite-dimensional Hilbert space structures such as the quantum combs formalism \cite{chiribella2008transforming} that introduced the highly studied quantum switch \cite{chiribella_quantum_2013}. Another prominent formalism from the perspective of post-quantum theories based on finite-dimensional Hilbert space structures is the process matrix formalism \cite{oreshkov2012quantum} that first studied bipartite indefinite causal structures that violate {\it causal inequalities}. Such directions have paved the way for understanding what quantum causality could look like through examples such as the relativistic realisations of the quantum switch \cite{belenchia2018quantum, moller2021quantum} as well as exploring the resource theoretical applications it may have \cite{chiribella2021indefinite,quintino2019probabilistic}. More recently, there have also been attempts to generalise quantum supermaps to process theories \cite{wilson2022mathematical}. In Fig.\ \ref{fig:GPTplusICS}, we present the schematic of where the Causaloid Framework may be placed in context to other works in the relevant fields.

The time is ripe to revisit the Causaloid framework and what it has to offer today in the context of recent literature in the field of study of indefinite causal structure. In this work\footnote{Based on Chapter 4 of the PhD thesis \cite{sakharwade2022operational}}, we present a diagrammatic language for the Causaloid framework and some natural new nomenclature for physical compression to facilitate exposition of the Causaloid framework, and by doing so provide a standalone review of the Causaloid framework through this new diagrammatic lens. The diagrammatic language presented here provides the necessary first steps to study in further detail the Causaloid framework to characterise the landscape of theories through Meta compression and find a hierarchy of theories thereof \cite{sakharwade2023hierarchy}. The diagrammatic language will also help us incorporate a family of GPTs that can be expressed as circuits (known as the Duotensors \cite{hardy2012operator,hardy2013formalism}), into the Causaloid framework helping situate it within the hierarchy found through Meta compression.

\subsection{Outline}

This diagrammatic review follows the papers that introduced the Causaloid framework \cite{hardy2005probability, hardy2007towards}, where \cite{hardy2007towards} is a compact version of longer \cite{hardy2005probability}. While this work can be read by itself, nonetheless we encourage referring to the papers \cite{hardy2005probability, hardy2007towards} liberally for the best comprehension experience. To this end, in Sections of this work, one is pointed to the corresponding relevant Sections from the papers \cite{hardy2005probability, hardy2007towards}, in case the reader wishes to supplement their reading. 

The work is structured in the following way. In Section \ref{sec:whycausaloid} we will discuss both the original and newer motivations for taking the Causaloid approach. In Section \ref{sec:setupcausaloid}, we first review the setup of the Causaloid framework: operational data, operational methodology (the person inside the box), how regions, generalised preparations and measurements are defined and lastly what we are interested in — predicting certain conditional probabilities; this portion is a condensed version based on Sections 2, 3, 5, 12-15 of \cite{hardy2005probability} and Section 3 of \cite{hardy2007towards}. In the subsequent Section \ref{sec:threelevelcompression} we review the fleshed-out framework through the three levels of physical compression — namely Tomographic, Compositional and Meta Compression; where we introduce a new diagrammatic representation for physical compression (building upon Sections 16-19 of \cite{hardy2005probability} and Section 4 of \cite{hardy2007towards}). The new diagrammatics will become quite useful in our following works \cite{sakharwade2023hierarchy, sakharwade2023duotensors}. We end this work with a succinct synopsis of the three levels of physical compression. In the final Section \ref{sec:causaloiddiscussion} we discuss the implications of our results and provide some exciting directions in which one may extend this work. Subsequent work will build on the diagrammatics introduced here, where we will study Meta Compression through the sufficiency of $d$-region Compositional compression, that defines a hierarchy of physical theories \cite{sakharwade2022operational,sakharwade2023hierarchy}.
 
\section{\label{sec:whycausaloid}Why the Causaloid Approach?}
In this Section, we briefly discuss the original motivations behind setting up the Causaloid framework \cite{hardy2005probability, hardy2007towards} as well as our motivations for revisiting this approach. 

\subsection{Towards Quantum Gravity...}
A theory of Quantum Gravity would reduce to General Relativity on the one hand and Quantum Theory on the other under the appropriate limits. General Relativity is a theory in which causal structure is dynamical, and the notion of simultaneity is operationally meaningless for space-like separated events. In Quantum Theory, any quantity that is dynamical is subject to quantum indefiniteness (for example, if a particle can go through one slit or another, then it can be indefinite as to which slit it goes through). It follows that we would expect \emph{indefinite causal structure} in a theory of Quantum Gravity. 

The question arises how one may find a mathematical basis for such a theory of Quantum Gravity. An even-handed approach is desired given the conceptual novelties brought in by both Quantum Theory (probabilistic nature) and General Relativity (dynamical causality) that are difficult to accommodate in the mathematical structures of the other. Such an even-handed approach is possible if one steps outside the confines of the mathematical formulations of either theory. The Causaloid framework \cite{hardy2005probability, hardy2007towards,hardy2009qgcomputers} is proposed as a general framework to accommodate them. 

\subsection{Letting go of states evolving in time}

In the standard formulation of QT, we have a state defined on a space-like hypersurface which we evolve with respect to an external time.  In contrast, in GR, time is introduced through the metric living on a four-dimensional manifold and its behaviour is dynamically determined by the field equations of the theory. We can, in fact, formulate GR in terms of a state defined on a surface evolving in time \cite{arnowitt1959dynamical}. However, this goes against the elegance of Einstein's manifestly covariant approach. Aside from this inelegance, there is another reason not to invoke an external time. In GR the causal structure is dynamically determined. In QT dynamically determined quantities are subject to quantum indefiniteness. This will likely mean that there is no matter of the fact as to whether a particular interval is spacelike or timelike. In other words, we may have an indefinite causal structure. In this case, it becomes impossible to foliate into spacelike hypersurfaces. Hence, we have to drop the idea of a state across space evolving in time.  
 
Therefore the Causaloid framework without assumptions on causal structure provides quite a general framework. We will see that the (in)definite causal structure is encoded in the second level of compression or Compositional Compression.  Note that if a theory has a definite causal structure, it is still instructive to formulate it without reference to a state evolving in time.  This is done, for example, in the duotensor framework \cite{hardy2010duotenzor}.  

\subsection{Operational Methodology, Data and Probabilities}

If the generality of the Causaloid framework stems from stripping us off \emph{a priori} assumptions of causality and the ontology of space-time is left open, then one may ask what lends us a stable footing that we rely on within this framework? The Causaloid framework is based on the assertion that \emph{any physical theory, whatever it does, must correlate recorded data}. 

This falls under the purview of operational methodology, wherein we find a pragmatic meeting point between differing philosophical positions. Agnostic of a physicist's philosophical tendencies towards (broadly speaking) realism and anti-realism within physics, we expect that a physical theory will \emph{at least} be consistent with experiments that produce certain classical data - such as the setup and observations recorded, say, on a piece of paper. The classical data will have correlations given the underlying physical theory and the goal is to then focus on systematically understanding the correlations within this data. These correlations allow us to compress the recorded data. 

Given the difficulties around finding a mathematical framework where Quantum theory and General Relativity meet, let alone finding a coherent ontology between the two, adopting an operational methodology as a basis is a safe route, perhaps even a desperate attempt to revisit the blank canvas before clouding it with assumptions that may not hold. The upside is that whatever we can recover operationally won't be wrong even if incomplete, and hopefully would lead towards a point where we are ready to tack on an ontology, should it be possible and desirable. 

It is important to note the distinction between using operationalism as a methodology and treating it as a philosophical viewpoint wherein physical concepts only have meaning with respect to some operational description.  Operational methodology is, perhaps, opportunistic but has been established as an invaluable tool in theoretical physics (such as in the development of Special Relativity).  Operationalism as a philosophical point of view, on the other hand, faces serious problems and has been much criticised in the philosophical literature \cite{sep-operationalism}.  

What a physical theory does among other things is help us predict quantities, in particular outcome probabilities, and the Causaloid shares this objective. Therefore, the mathematics and probability theory will inform how compression works. Hence the Causaloid framework can be regarded as a ``general probability theory for theories with indefinite causal structure''; it may also be regarded as a ``theory to study correlated data''. 

\subsection{Diagrammatics: Old Framework, New Clothes}

This brings us to the point where we address why we decided to revisit this framework. The Causaloid framework, by providing key insights and motivations, has been an important contribution guiding the study of \emph{indefinite causal structures}, often cited for such reasons; and yet, there has been only a handful of papers (for example \cite{markes2011entropy}) specifically building upon the Causaloid framework. This is due to two main reasons.

Firstly, since the introduction of the Causaloid framework, other operational frameworks have come up that can be used to study indefinite causality, such as Quantum Combs \cite{chiribella2009theoretical}, Process Matrices \cite{oreshkov2012quantum} and Causal Boxes \cite{portmann2017causal} to name a few. These frameworks are constructed with some assumptions that aid in their direct applicability where they can be seen as generalisations of quantum information processes. This feature leads to the possibility of more specific results that are harder to achieve in the more general Causaloid framework. Though, depending on the goal in mind a bug can become a feature, and the generality of the Causaloid serves the purpose of being able to accommodate any theory that studies correlated data. To this end, in this work we study further the third level of compression --- Meta Compression, to find any broad statements that can be made for physical theories, and in fact it will provide us with a way to categorise theories into a hierarchy (this is explored in our second paper \cite{sakharwade2023hierarchy} of this trilogy).     

Secondly, the Causaloid framework is quite abstract requiring the introduction of its terminology.  We present a diagrammatic representation of the framework that will perhaps make it somewhat easier for the interested person to be able to understand and work with the Causaloid framework. The diagrammatics will prove to be quite powerful, we will see how the \emph{duotensors} \cite{hardy2011reformulating} relate to the Causaloid framework, primarily through diagrams.

\section{\label{sec:setupcausaloid}Causaloid framework}
In this Section, we cover the setup of the Causaloid framework: recording operational data, the person inside the box inferring the underlying physical theory through an operational methodology, how regions are defined and lastly what we are interested in — predicting conditional probabilities. Note that this portion is based on material from Sections 2, 3, 5, 12-15 of \cite{hardy2005probability} and Section 3 of \cite{hardy2007towards}.

\subsection{Thinking inside the box}

Now we wish to describe how the Causaloid framework is set up using an operational methodology wherein data is recorded onto cards.  First, we will provide a few examples to illustrate the idea.  

\begin{figure}[h!]
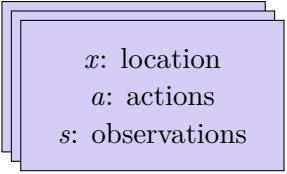

\begin{center}
\begin{Diagram}{3}{-20}
\begin{move}{0,0}

\fill[blue!80!red!20!](-8,1) -- (-8,9) -- (6,9) -- (6,1) -- (-8,1);
\draw (-8,1) -- (-8,9) -- (6,9) -- (6,1) -- (-8,1);
\fill[blue!80!red!20!](-7.5,0.5) -- (-7.5,8.5) -- (6.5,8.5) -- (6.5,0.5) -- (-7.5,0.5);
\draw (-7.5,0.5) -- (-7.5,8.5) -- (6.5,8.5) -- (6.5,0.5) -- (-7.5,0.5);
\fill[blue!80!red!20!](-7,0) -- (-7,8) -- (7,8) -- (7,0) -- (-7,0);
\draw (-7,0) -- (-7,8) -- (7,8) -- (7,0) -- (-7,0);
\draw (0,6) node {\textit{x}: location};
\draw (0,4) node {\textit{a}: actions};
\draw (0,2) node {\textit{s}: observations};
\end{move}
\end{Diagram}
\caption{\label{fig:cards} Data recorded on Cards}
\end{center}
\end{figure}

Consider the idea of a run of an experiment.  An experiment generally takes place over some period of time and over some region of space.  For example, we could imagine several qubits passing through a circuit where each operation in the circuit in the circuit has a setting and an outcome.  We can label each operation with a number which we can call `` location\rq\rq. Before running the experiment we need to choose the settings.  We can do this by specifying the setting as a function of the location.  We call this specification a \emph{procedure}. To initiate this experiment we would release the qubits (perhaps by pressing a button on the operations at the start of the circuit).  Then, at each operation, we can collect information consisting of (location, setting, outcome). We will imagine recording the information collected at each operation on a card (as shown in Fig.\ \ref{fig:cards}).  At the end of the run of the experiment, we will have collected a stack of cards (one card from each operation). We imagine bundling this stack of cards and labelling it with a tag that specifies the procedure.  We can run this experiment many times with the same procedure so we can collect probabilistic information.  Then run it many times with every other possible procedure.  After each run, we collect a tagged bundled stack of cards.  Collectively all these stacks provide exhaustive information about this experiment.   

We could consider a different kind of example where, rather than having operations fixed in a circuit arrangement, we have some probes floating in space which interact with particles in the same region of space. Each probe has a GPS tracker to read off location, some knobs to choose settings, and some detectors whose clicks represent outcomes.  Further, we imagine that each probe has a \lq\lq tick time\rq\rq\: and during each tick, data is outputted onto a card representing (location, setting, and outcome).   We specify a procedure by specifying a function from location (i.e.\ GPS reading in this example) to setting.  We can initiate the experiment (perhaps by releasing the probes from some given configuration) and allow each probe to operate for some given number of ticks.  After one run of the experiment, we will collect a stack of cards (one card from each probe for each tick).  We can tag, bundle and repeat as before.  In this example, the set of locations we expect to record on the cards are not fixed in advance and may vary from run to run.  

We can apply this idea to pretty much any experiment (certainly one where we record proximate chunks of data like this).  Thus, in general, during a run of an experiment, we will collect data from different parts of the experiment.   We think of recording this data onto some number of cards (as shown in Fig.\ \ref{fig:cards}).  
Each card contains some \emph{proximate} data (which we imagine as being collected in some small vicinity) corresponding to location, actions (choice of settings), and outcomes. Since this data is classical it could instead, of course, be written on a computer or stored in other ways, nonetheless, we may use the idea of cards to setup this framework for the purpose of exposition.  We need a way to choose the settings at each location.  We do this by defining a \emph{procedure}.  
A procedure is specified by letting the action be given as a function of location. Given such a procedure, we run the experiment once and collect a stack of cards. We bundle this stack of cards and attach a tag with the procedure used. Then we repeat the experiment many times, with this procedure, collecting a bundled tagged stack of cards each time. This allows us to extract probabilistic information. Then we can repeat the experiment with all other possible procedures. This generates many additional bundled and tagged stacks of cards. All these stacks, taken together, contain all the logical operational information available by this process. 

Now, imagine a person inside a closed room, having access to all these bundled and tagged stacks of cards. This person is tasked with inferring (aspects of) the underlying physical theory that governs the data. The recorded data will exhibit correlations due to this underlying physical theory. This correlation implies a kind of redundancy. The person in the box distils away the redundancy by \emph{compressing} the data as we will describe. 
 
The person is not able to look outside the box for extra information and must therefore define all concepts using only the information provided through the cards. This enforces a certain kind of operational honesty\footnote{Note that there may not be an actual person in a box, rather this \emph{gedanken} setup is to serve the purpose of forcing us to adopt an operational methodology.}. 

\subsection{\label{sec:cards}Organising Stacks of Cards}

Each card records a small amount of proximate data. We may think of the cards as representing something analogous to space-time events. One piece of data recorded on any given card will be something we will regard as representing or being analogous to space-time location.

We will consider examples where the data recorded on each card is sorted in the form --- (\textit{x - Location}, \textit{a - Actions}, \textit{s - Observation}) (or in short (\textit{x,a,s})), where each of the pieces of information are categorised as follows:

\begin{enumerate}
    \item \textbf{Location:} The first piece of data, \textit{x}, is an observation and represents location. It could be some real physical reference frame such as a GPS system or it could be some other data that we are simply going to regard as representing location.
    \item \textbf{Actions:} The second piece of data, \textit{a}, represents some actions. For example, it might correspond to the configuration of some knobs we have freedom in setting.
    \item \textbf{Observations:} The third piece of data, \textit{s}, represents some further local observations, such as the outcome that is obtained. 
\end{enumerate} 
Given this form of recorded data, we can organise the cards into the following sets. 
\begin{itemize}
    \item\textbf{The stack}, denoted by $Y$, is the set of cards from a single run of an experiment.
    \item\textbf{The procedure}, denoted by $F$, corresponds to all cards that are consistent with the given function of the settings, where we rewrite actions $a$ as a function $F(x)$ depending on location $x$ (thus the data is of the form (\textit{x, F(x), s})). $F$ here stands for the procedure set of cards, the Function $F(x)$ as well as the procedure given by the description of actions, and it will be clear from context which meaning is implied.
    \item\textbf{The full pack}, denoted by $V$, is the set of all logically possible cards when all possible locations, procedures and outcomes are taken into account. It is possible that some cards never actually occur in any stack because of the nature of the physical theory but they are included anyway.
    \textbf{}

\end{itemize}
 The procedure $F$ can be thought of as ``what was done\rq\rq. The stacks $Y$ (particular run of procedure giving some outcome) can be thought of as ``what was seen\rq\rq.   

Given the definitions of the sets we have the following relations between them:
\begin{align}
    Y \quad\textit{(Outcome \big| Procedure)} \quad\subseteq\quad F \quad\textit{(Procedure)} \quad\subseteq\quad V \quad\textit{(Full pack)}
\end{align}

\subsection{\label{sec:regions}Operational Regions}

The set $V$ is the complete set of logical possible cards (the stacks). We discussed the subsets of $V$ with respect to \textit{a} - Actions (given by subset Procedure $F$), and with respect to \textit{s} - Observations (given by subset Outcome $Y$); now we look at subsets with respect to \textit{x} - Locations which we call \emph{regions}. The region $R_O$ is specified by the set of cards from $V$ having $x\in O$. We define $R_x$ to be an elementary region consisting only of the cards having $x$ on them.

\begin{figure}[h!]
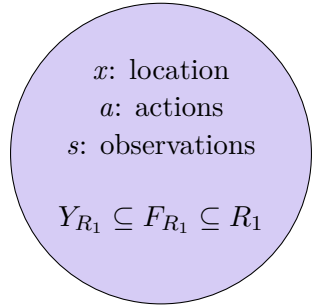

\begin{center}
\begin{Diagram}{3}{-20}
\begin{move}{0,0}
\fill[blue!80!red!20!] (0,1.5) circle (8);
\draw (0,1.5) circle (8); 
\draw (0,6) node {\textit{x}: location};
\draw (0,4) node {\textit{a}: actions};
\draw (0,2) node {\textit{s}: observations};
\draw (0,-2) node {$Y_{R_1} \subseteq F_{R_1} \subseteq R_1$};
\end{move}
\end{Diagram}
\caption{\label{fig:op_region} Operational Region $R_1$}
\end{center}
\end{figure}

Within regions, there can be many possible sets of actions (associated with procedures $F$, $F'$,...). In a region, we consider having an independent choice of which action (procedure) to implement. This captures the notion of ``space-time\rq\rq\footnote{Here, we use the term space-time generally. It may be seen as some general notion of spatio-temporal regions (perhaps indefinite) that may be distinct from the relativistic notion of space-time.} regions where we have local choices. When we have a particular run of the experiment given some procedure we end up with a stack $Y$ of data that can be sorted by regions. Then we find a picture of what happened, laid out in a kind of ``space-time\rq\rq\: or more precisely in terms of operational regions.

For a region $R_1$ (shorthand for $R_{O_1}$) we define the outcomes in $R_1$ ($Y_{R_1}$) as the cards from the stack $Y$ that are attributed to region $R_1$: 
\begin{align}
   Y_{R_1} = Y \cap R_1 
\end{align}
Similarly, we define procedure in $R_1$ ($F_{R_1}$) as the cards from the set $F$ that are attributed to the region $R_1$
\begin{align}
    F_{R_1} = F \cap R_1
\end{align}
Clearly $Y_{R_1} \subseteq F_{R_1} \subseteq R_1$. Given ($Y_{R_1}$, $F_{R_1}$) we know “what was done” ($F_{R_1}$) and “what was seen” ($Y_{R_1}$) in region $R_1$.

\subsection{\label{subsec:welldefinedprobabilities}Statement of objective}

The objective of the Causaloid framework is to predict (when possible) the conditional probabilities of outcomes given the procedure as well as the outcomes and procedures in different disjoint regions.  These conditional probabilities take the form
\begin{align}
    {\rm prob}(Y_{R_1}| Y_{R_2}, ..., Y_{R_n}, F_{R_1}, F_{R_2}, ... , F_{R_n})
\end{align}
without any assumptions of the causal structure, given that $R_i$, $i \in \{1,2,...,n\}$ are $n$ disjoint regions. Operationally, one may interpret the probability as the relative frequencies of the number of cards 
\begin{align}
   \frac{N(Y_{R_1},Y_{R_2},...,Y_{R_n},F_{R_1},...,F_{R_n})}{N(Y_{R_2},...,Y_{R_n},F_{R_1},...,F_{R_n})} \label{eqn:ratiofreqprob}
\end{align}
when $N(\cdot)$ (the number of cards of the type given in parenthesis) becomes large, at least for the denominator.

An important question arises: to begin with are these probabilities even well-defined and if yes, how can we use the Causaloid framework to efficiently calculate this? Let us address the first part of the question here; the next Section is dedicated to answering the latter. To understand the issue around well-defined probabilities consider three consecutive polarisers (associated with three regions) placed at some angle specified as part of the procedure of their respective regions, the outcome notes if a photon passes through or is absorbed. Then the probability of the outcome of the third polariser conditioned on the outcome of the first, as well as the procedures of both of them, given by,
\begin{align}
    {\rm prob}(Y_{\text{third polariser}}| Y_{\text{first polariser}}, F_{\text{first polariser}}, F_{\text{third polariser}})
\end{align}
is not well defined without consideration of the second polariser and the framework will not be able to predict any value for this conditional probability.  On the other hand, the probability of the outcome for the second polariser, conditioned on the outcomes of the first and third, as well as the procedures on all three polarisers, given by,
\begin{align}
    {\rm prob}(Y_{\text{second polariser}}| Y_{\text{first polariser}}, Y_{\text{third polariser}}, F_{\text{first polariser}},F_{\text{second polariser}}, F_{\text{third polariser}})
\end{align}
is well-defined.  The difference between these two scenarios has to do with how the proposition, whose probability we are interested in calculating, relates to the background causal structure assumed in this polariser example.  

Thus, when making predictions, we require the following two-step process. First, we do prediction heralding: is the probability we are interested in calculating well-defined in the first place?  Second, if this probability is well defined, then what is it equal to?  This two-step process seems inevitable if we have an indefinite causal structure since then we cannot rely on a fixed background causal structure to provide intuition as to which probabilities are well defined. After setting up the Causaloid framework, we will discuss in Section \ref{sec:calculatingprobabilities} how to do prediction heralding and how to calculate probabilities when the heralding tells us they are well-defined.  

To get started, we restrict ourselves to a large region $R$ that would contain a large number of (but not all) cards from the full pack $V$ for which all probabilities 
\begin{align}{\rm prob}(Y_R|F_R,C)\end{align} 
are well-defined given some conditioning $C$ on cards outside $R$. In the context of the example of the polariser the region $R$ may be considered to be the room in which the experiment takes place and the conditioning $C$ may be all the requirements to avoid ``noise\rq\rq\: in the experiments. At the level of theory encompassing everything this conditioning may be argued either to take the form of ``boundary conditions\rq\rq\: (where $R$ is \emph{open} and associated with the ``Open Causaloid\rq\rq, see Section 22 of \cite{hardy2005probability}) or not be required (where $R$ is \emph{closed} and associated with the``Universal Causaloid\rq\rq, see Section 30 and 35 of \cite{hardy2005probability}) 

Going forward we always consider this large region $R$ which is called a ``\emph{predictively well-defined region}\rq\rq\: and we simply write 
\begin{align}
    {\rm prob}(Y_R|F_R)
\end{align}
where the conditioning on $C$ is taken to be implicit.  

\section{\label{sec:threelevelcompression}Three levels of Physical Compression of Data}

In this Section, we review how to distil the redundancy in recorded data (cards) through three levels of physical compression to form a ``theory of correlated data''. By compression, we mean the process of organising the data recorded on the cards, and reducing the set of data to the minimum; where the minimum required data can be used to predict any desired probabilities. We call it physical compression since it is contingent on the physical theory that correlates the data. The levels of compression are as follows:

\begin{quote}
    \textbf{Pre-Compression:} The zeroth level of compression introduces the notion of generalised states and preparations, in the absence of assumptions of causal relations between regions, and sets up the stage for the following levels. 

    \textbf{Tomographic Compression:} The first level of compression pertains to physical compression over a single region and here the compression matrices ($\Lambda$) and compression sets ($\Omega$) are introduced. Fiducial states are also introduced. This level is closely related to the concept of \emph{tomography} and its relation to Generalised Probability Theories (GPTs) will be discussed.
    
    \textbf{Compositional Compression:} The second level of compression pertains to physical compression over disjoint regions through their \emph{composition}. Here, the notion of \emph{causal adjacency} is introduced. It captures strong causal connections between regions when non-trivial Compositional Compression occurs.
    
    \textbf{Meta Compression:} The third and final level pertains to compression over Tomographic and Compositional Compression, since there may be some universal, region independent \emph{rules} related to the mathematical structure of the theory, which gives the Causaloid ${\bf \Lambda}$, seen as a specification of the physical theory itself.  Here, the \emph{Causaloid product} $\otimes^{\Lambda}$ that generalises temporal and spatial products is introduced. This is used to take the product over regions and is the culmination of the compression process.  
\end{quote}
This review of the levels of compression closely follows concepts explained in Sections 16-19 of \cite{hardy2005probability} and Section 4 of \cite{hardy2007towards}; additionally, here we present a new diagrammatic representation of the Causaloid framework. The terminology of three levels of compression --- Tomographic, Compositional and Meta (previously first, second and third level) are introduced in this work to aid conceptual understanding of these compression levels. 

\subsection{\label{precompression}Zeroth level: Pre-Compression}
Recall the classical data recorded on cards in an operational manner. A physical theory will capture the correlations in this data and express them through physical concepts. While some of these physical concepts may be theory-dependent, the concept of a physical state is central to most (if not all) physical theories. Before we discuss the three levels of compression, we must focus on how we may represent states in this framework with the data. The definition of states is often either based on a notion of preparation that happens in the past to give the state we have in the present; or based on a notion of measurement that happens in the future of the state that the measurement outcomes tell us something about. Thus these rely on a causal structure that is assumed to be fixed, in so far as the order of events being such that preparation happens in the past and measurement happens in the future. Since we are interested in the absence of assumptions on the underlying causal structure between elementary regions, we need a general notion of physical state, preparations and measurements.  A (generalised) state for a region associated with a (generalised) preparation outside of that region is taken to be that thing given by any mathematical object that can be used to calculate the probability for all outcomes of all measurements that might be performed in that region.  

Of course, the above definition is minimal in that it centres on the operational role of a state while being agnostic of the ontological status of the state. This is ideal since we are building up from operational data. Let us now apply this notion of state to the stacks of cards. Consider a region $R_1$ inside the predictively well-defined region $R$ (note that $R_1$ need not be an elementary region). Since $R = R_1\cup (R-R_1)$ we can write:
\begin{align}\label{split}
p &={\rm prob}(Y_{R}| F_{R})\\
  &={\rm prob}(Y_{R_1}\cup Y_{R-R_1}| F_{R_1}\cup F_{R-R_1})
\end{align}
We can regard $(Y_{R-R_1}, F_{R-R_1})$ which happens in $R-R_1$ as a {\it generalised preparation} for region
$R_1$. The generalised preparation for $R_1$ is considered to ``prepare\rq\rq\: a state which is defined soon below. Further, we can regard $(Y_{R_1}, F_{R_1})$ which happens in $R_1$ as a {\it measurement plus outcome} associated with the actions and outcomes seen in $R_1$ (the choice of measurement is $F_{R_1}$ and the outcome of the measurement is $Y_{R_1}$).   It is useful to give each such {\it measurement plus outcome} in the region $R_1$ a label, $\alpha_1$, where $\alpha_1$ belongs to the (possibly very big) set $\Gamma_{R_1}$.  That is we have:
\begin{align}
 (Y_{R-R_1}, F_{R-R_1}) \Longleftrightarrow& \textit{generalised preparation for $R_1$} \label{YRprep}\\
 (Y^{\alpha_1}_{R_1}, F^{\alpha_1}_{R_1}) \Longleftrightarrow& \textit{measurement in $R_1$} \label{YRmeasure}
\end{align}
Then we can also label the probabilities with $\alpha_1$ denoting the probabilities associated with measurement outcomes labelled $\alpha_1$ as follows:
\begin{align}
    p_{\alpha_1}={\rm prob}(Y^{\alpha_1}_{R_1}\cup Y_{R-R_1}&| F^{\alpha_1}_{R_1}\cup F_{R-R_1})\label{splitalpha} =
\begin{DiagramV}{3}{0}
\begin{move}{0,0}
\draw (0,0) -- (6,0); 
\fill[blue!80!red!20!] (6,0) circle (2);
\draw (6,0) circle (2) node {$p$};
\draw (2.5,1) node {\footnotesize $\alpha_{1}$};
\end{move}
\end{DiagramV}
\end{align}
We will introduce the diagrammatic representation alongside the mathematical definition, similar to Equation \eqref{splitalpha} throughout this Section.

Note that in Equation \eqref{splitalpha} above, we choose to consider joint probabilities over outcomes rather than the conditional probability ${\rm prob}(Y_{R_1}| Y_{R-R_1}, F_{R_1}\cup F_{R-R_1})$ which we might desire.  This is intentional since conditional probabilities introduce nonlinearities (since they are obtained by dividing by a joint probability).  It turns out that these nonlinearities become especially problematic when we have an indefinite causal structure. One can always calculate conditional probabilities at the end of the calculation when required. 

The definition of the state in the context of the Causaloid framework (taken from \cite{hardy2005probability}) is as follows:
\begin{quote}
{\bf The state (in Causaloid framework)} for $R_1$ associated with a generalised preparation in $R-R_1$ is defined to be the thing represented by any mathematical object which can be used to predict $p_{\alpha_1}$ for all measurements in $R_1$ labelled by the index ${\alpha_1}$.
\end{quote}
We could simply write the state ${\bf P}(R_1)$ as a vector that lists all the probabilities $p_{\alpha_1}$ since it satisfies the above definition:
\begin{equation}
{\bf P}(R_1)= \left( \begin{array}{c} \vdots \\ p_{\alpha_1} \\ \vdots \end{array} \right)\label{eqnStateAlpha}=
\begin{DiagramV}{3}{-0.5}
\begin{move}{0,0}
\draw (6,6) -- (6,0); 
\fill[black] (6,4) circle (0.25);
\fill[blue!80!red!20!] (6,0) circle (2);
\draw (6,0) circle (2) node {${\bf P}$};
\end{move}
\end{DiagramV}{\rm ~~where~~} \alpha_1 \in \Gamma_{R_1}
\end{equation}
where $\alpha_1$ belongs to the set $\Gamma_{R_1}$ (or simply $\Gamma_{1}$) which labels all actions and outcomes and we expect the size of the set $\Gamma_{1}$ to be quite big (and possibly infinite).

The objective is to be able to recover the probabilities $p_{\alpha_1}$ and to do so we define another vector ${\bf R}_{\alpha_1}$ which represents the measurement $(Y^{\alpha_1}_{R_1}, F^{\alpha_1}_{R_1})$ in region $R_1$ such that
\begin{align}
p_{\alpha_1}&= {\bf R}_{\alpha_1}(R_1)\cdot {\bf P}(R_1)\label{eqnPR}\\
\begin{DiagramV}{0}{0}
\begin{move}{0,0}
\draw (0,0) -- (6,0); 
\fill[blue!80!red!20!] (6,0) circle (2);
\draw (6,0) circle (2) node {$p$};
\draw (2.5,1) node {\footnotesize $\alpha_{1}$};
\end{move}
\end{DiagramV}~~~&=~~~
\begin{DiagramV}{3}{+0.5}
\begin{move}{0,0}
\draw (0,0) -- (6,0); 
\draw (2.5,1) node {\footnotesize $\alpha_{1}$};
\draw (6,-5) -- (6,-2);
\fill[black] (6,-3.5) circle (0.25);
\fill[blue!80!red!20!] (6,0) circle (2);
\draw (6,0) circle (2) node {${\bf R}$};
\fill[blue!80!red!20!] (6,-7) circle (2);
\draw (6,-7) circle (2) node {${\bf P}$};
\end{move}
\end{DiagramV}\label{diaPR}
\end{align}
where the vector ${\bf R}_{\alpha_1}(R_1)$ has 1 in position ${\alpha_1}$ and 0's in all other positions. The inner product of the two vectors gives us $p_{\alpha_1}$ and the above equation may be thought of being some generalisation of the Born rule, under the generalised notions of states ${\bf P}(R_1)$ and measurements ${\bf R}(R_1)$ given above. Note that Equation \eqref{eqnPR}-\eqref{diaPR} is linear in ${\bf P}(R_1)$ and ${\bf R}(R_1)$ and that in Equation \eqref{diaPR} the line with a filled black dot represents the inner product.

\subsection{\label{tomographiccompression}First level: Tomographic Compression}

Now we are prepared to tackle the first level of compression. Here we continue considering a single region $R_1$. While through Equation \eqref{eqnStateAlpha} we were able to calculate (well defined) probabilities $p_{\alpha}$ in terms of the generalised preparation of state for $R_1$ and measurements in $R_1$, notice that the vector ${\bf P}(R_1)$ can be quite long and the set $\Gamma_{R_1}$ can be quite big. A physical theory will (we expect) have some structure for state space that allow for the specification for fewer probabilities (or in other words fewer procedures and outcomes) to be able to predict any $p_{\alpha_1}$. For example, we encounter this in Quantum Theory through the process of \emph{Tomography}. 

Therefore we may replace ${\bf P}(R_1)$ with a list of the smallest subset of probabilities from $p_{\alpha_1}$ that allows us to predict any $p_{\alpha_1}$ by means of linear relations. To do so we pick a set of {\it fiducial} measurements $(Y^{l_1}_{R_1}, F^{l_1}_{R_1})$ labelled by $l_1 \in \Omega_{R_1}$ (or simply $\Omega_1$) where $\Omega_{R_1} \subseteq \Gamma_{R_1}$ and the (new representation of the) state (compared to Equation \eqref{eqnStateAlpha}) becomes:

\begin{equation}
{\bf p}(R_1)= \left( \begin{array}{c} \vdots \\ p_{l_1} \\ \vdots \end{array} \right)=
\begin{DiagramV}{3}{-0.5}
\begin{move}{0,0}
\draw (6,6) -- (6,0); 
\fill[black] (6,4) circle (0.25);
\fill[blue!80!red!20!] (6,0) circle (2);
\draw (6,0) circle (2) node {${\bf p}$};
\end{move}
\end{DiagramV} {\rm ~~where~~}
l_1\in\Omega_{R_1}
\end{equation}

Note that the choice of the set of fiducial measurements need not be unique, and one may pick any one such set but the size of the set $\Omega_{R_1}$ (or the length of ${\bf p}(R_1)$) will be the same over any such choice which is the minimum possible size such that ${\bf p}(R_1)$ continues to satisfy the definition of a state. The sets $\Omega_{R_O}$ (for some region $R_O$) play an important role in the Causaloid framework and its size will be informed by the system and underlying physical theory. Of course, if no such subset can be found we can always resort to having ${\bf p}(R_1)={\bf P}(R_1)$ and $\Omega_{R_1} = \Gamma_{R_1}$ (which will give no compression).

Now that we have the compressed state ${\bf p}(R_1)$ we need a way to calculate any general $p_{\alpha}$ from it and require a \textit{linear} formula analogous to Equation \eqref{eqnPR}. Let us express this as:

\begin{align}\label{rdotp}
p_{\alpha_1}&= {\bf r}_{\alpha_1}(R_1)\cdot {\bf p}(R_1)\\
\begin{DiagramV}{0}{0}
\begin{move}{0,0}
\draw (0,0) -- (6,0); 
\fill[blue!80!red!20!] (6,0) circle (2);
\draw (6,0) circle (2) node {$p$};
\draw (2.5,1) node {\footnotesize $\alpha_{1}$};
\end{move}
\end{DiagramV}~~~&=~~~
\begin{DiagramV}{3}{+0.5}
\begin{move}{0,0}
\draw (0,0) -- (6,0); 
\draw (2.5,1) node {\footnotesize $\alpha_{1}$};
\draw (6,-5) -- (6,-2);
\fill[black] (6,-3.5) circle (0.25);
\fill[blue!80!red!20!] (6,0) circle (2);
\draw (6,0) circle (2) node {${\bf r}$};
\fill[blue!80!red!20!] (6,-7) circle (2);
\draw (6,-7) circle (2) node {${\bf p}$};
\end{move}
\end{DiagramV}
\end{align}
The exact form of ${\bf r}_{\alpha_1}(R_1)$ can be found from the recorded data and will contain real numbers, nonetheless we can provide the form for some of these. Given the fiducial measurements $(Y^{l_1}_{R_1}, F^{l_1}_{R_1})$ in $R_1$, ${\bf r}_{\alpha_1}(R_1)$ for $\alpha_1 = l_1$ would simply be 
\begin{equation}\label{fidmeas}
{\bf r}_{l_1} =\left( \begin{array}{c} 0 \\ 0 \\ \vdots \\ 1 \\ \vdots \\ 0 \end{array} \right) {\rm ~~for~all~~}
{l_1}\in\Omega_{R_1}
\end{equation}
where 1 is in the ${l_1}^{th}$ position and all other entries are 0's since this is the only way to ensure $p_{l_1}= {\bf r}_{l_1}(R_1)\cdot {\bf p}(R_1)$. The fiducial measurements thus form a basis.  

Now we can see that the general vector ${\bf r}_{\alpha_1}(R_1)$ can be expressed as a linear combination over the basis of measurements spanned by ${\bf r}_{l_1}(R_1)$. This linear relation can be defined by introducing the compression matrix $\Lambda_{\alpha_1}^{l_1}$ as follows:
\begin{equation}\label{firstlevellambdar}
{\bf r}_{\alpha_1}= \sum_{l_1\in\Omega_{R_1}} \Lambda_{\alpha_1}^{l_1} {\bf r}_{l_1}
\end{equation}

\begin{equation}
\begin{DiagramV}{0}{0}
\begin{move}{8,0}
\draw (15+5-20,0) -- (26-20,0); 
\draw (26-20,-5) -- (26-20,-2);
\fill[black] (26-20,-3.5) circle (0.25);
\fill[blue!80!red!20!] (26-20,0) circle (2);
\draw (26-20,0) circle (2) node {${\bf r}$};
\draw (22-20,1) node {\footnotesize $\alpha_{1}$};
\end{move}
\end{DiagramV}
=
\begin{DiagramV}{0}{0}
\begin{move}{0,0}
\draw (7-7,0) -- (11-7,0); 
\draw (15-7,0) -- (21-7,0); 
\draw (21-7,-5) -- (21-7,-2);
\fill[black] (21-7,-3.5) circle (0.25);
\fill[blue!80!red!20!] (21-7,0) circle (2);
\fill[blue!80!red!20!] (11-7,-2) rectangle (15-7,2);
\draw (21-7,0) circle (2) node {${\bf r}$};
\draw (11-7,-2) rectangle (15-7,2);
\draw (13-7,0) node {$\Lambda$};
\draw (9-7,1) node {\footnotesize $\alpha_{1}$};
\draw (17-7,1) node {\footnotesize $l_{1}$};
\end{move}
\end{DiagramV}
\end{equation}
We can take the inner product of Equation \eqref{firstlevellambdar} with some general ${\bf p}$ to find a similar relation for probabilities:
\begin{align}
&{\bf r}_{\alpha_1}\cdot{\bf p}= \sum_{l_1\in\Omega_{R_1}} \Lambda_{\alpha_1}^{l_1} {\bf r}_{l_1} \cdot{\bf p} \nonumber\\
\Rightarrow \quad& p_{\alpha_1}= \sum_{l_1\in\Omega_{R_1}} \Lambda_{\alpha_1}^{l_1} p_{l_1} \label{firstlevellambdap} 
\end{align}

\begin{equation}
\begin{DiagramV}{0}{0}
\begin{move}{0,0}
\draw (15+5-20,0) -- (26-20,0); 
\fill[blue!80!red!20!] (26-20,0) circle (2);
\draw (26-20,0) circle (2) node {${p}$};
\draw (22-20,1) node {\footnotesize $\alpha_{1}$};
\end{move}
\end{DiagramV}
=
\begin{DiagramV}{0}{0}
\begin{move}{0,0}
\draw (7-7,0) -- (11-7,0); 
\draw (15-7,0) -- (21-7,0); 
\fill[blue!80!red!20!] (21-7,0) circle (2);
\fill[blue!80!red!20!] (11-7,-2) rectangle (15-7,2);
\draw (21-7,0) circle (2) node {${p}$};
\draw (11-7,-2) rectangle (15-7,2);
\draw (13-7,0) node {$\Lambda$};
\draw (9-7,1) node {\footnotesize $\alpha_{1}$};
\draw (17-7,1) node {\footnotesize $l_{1}$};
\end{move}
\end{DiagramV}
\end{equation}
It is evident from the definition of the compression matrix $\Lambda_{\alpha_1}^{l_1}$ in Equation \eqref{firstlevellambdar} that its components can be fixed using the entries of ${\bf r}$-vectors:
\begin{align}\label{rfirstlevelcomponent}
{\bf r}_{\alpha_1}|_{l_1} = \Lambda^{l_1}_{\alpha_1} := \begin{DiagramV}{3}{0}
\begin{move}{0,0}
\draw (1,0) -- (11-0.1,0); 
\draw (15,0) -- (25,0); 
\fill[blue!80!red!20!] (11,-2) rectangle (15,2);
\draw (11,-2) rectangle (15,2);
\draw (13,0) node {$\Lambda$};
\draw (6,1) node {\footnotesize $\alpha_{R_1}\in \Gamma_{R_1}$};
\draw (20,1) node {\footnotesize $l_{R_1}\in \Omega_{R_1}$};
\end{move}
\end{DiagramV}
\end{align}
where ${\bf r}_{\alpha_1}|_{l_1}$ is the $l_1^{th}$ component of ${\bf r}_{\alpha_1}$.

Note that $\Lambda_{\alpha_1}^{l_1}$ will (often) be a very rectangular matrix since we expect that $|\Gamma_{R_1}| \geq |\Omega_{R_1}|$ where $|.|$ gives you the size of the sets. From the definition for the $\Lambda$ matrix we also have:
\begin{equation}\label{lambdadelta}
\Lambda_{l'_1}^{l_1} = \delta_{l'_1}^{l_1} {\rm ~~~for~~~} l'_1, l_1 \in \Omega_{R_1}
\end{equation}
where $\delta_{l'_1}^{l_1}$ equals 1 if the subscript and superscript are equal and is 0 otherwise.

This concludes the details for the first level compression. Let us recapitulate for assimilation. The person in the box in the pre-compression phase, first organises the recorded data by regions R, procedures $F$ and outcomes $Y$ and finds the state ${\bf P}$ associated with regions. By the structure of the underlying physical theory the state ${\bf P}$ can be \emph{compressed} to ${\bf p}$ and the measurements ${\bf R}_{\alpha_1}$ can be \emph{compressed} to ${\bf r}_{\alpha_1}$ where the relation of ${\bf r}_{\alpha_1}$ to ${\bf r}_{l_1}$ is captured by the $\Lambda$ matrix:
\begin{align}
p_{\alpha_1}&= {\bf R}_{\alpha_1}(R_1)\cdot {\bf P}(R_1)\\
&= {\bf r}_{\alpha_1}(R_1)\cdot {\bf p}(R_1)\\
&= \sum_{l_1\in\Omega_{R_1}} \Lambda_{\alpha_1}^{l_1} {\bf r}_{l_1}(R_1)\cdot {\bf p}(R_1)
\end{align}
or diagrammatically we have,
\begin{align}
\begin{DiagramV}{0}{0}
\begin{move}{0,0}
\draw (0,0) -- (6,0); 
\fill[blue!80!red!20!] (6,0) circle (2);
\draw (6,0) circle (2) node {$p$};
\draw (2.5,1) node {\footnotesize $\alpha_{1}$};
\end{move}
\end{DiagramV}=%
\begin{DiagramV}{3}{+0.5}
\begin{move}{0,0}
\draw (0,0) -- (6,0); 
\draw (2.5,1) node {\footnotesize $\alpha_{1}$};
\draw (6,-5) -- (6,-2);
\fill[black] (6,-3.5) circle (0.25);
\fill[blue!80!red!20!] (6,0) circle (2);
\draw (6,0) circle (2) node {${\bf R}$};
\fill[blue!80!red!20!] (6,-7) circle (2);
\draw (6,-7) circle (2) node {${\bf P}$};
\end{move}
\end{DiagramV}=%
\begin{DiagramV}{3}{+0.5}
\begin{move}{0,0}
\draw (0,0) -- (6,0); 
\draw (2.5,1) node {\footnotesize $\alpha_{1}$};
\draw (6,-5) -- (6,-2);
\fill[black] (6,-3.5) circle (0.25);
\fill[blue!80!red!20!] (6,0) circle (2);
\draw (6,0) circle (2) node {${\bf r}$};
\fill[blue!80!red!20!] (6,-7) circle (2);
\draw (6,-7) circle (2) node {${\bf p}$};
\end{move}
\end{DiagramV}=%
\begin{DiagramV}{3}{+0.5}
\begin{move}{0,0}
\draw (-8,0) -- (0,0); 
\fill[blue!80!red!20!] (-4,-2) rectangle (0,2);
\draw (-4,-2) rectangle (0,2);
\draw (-2,0) node {$\Lambda$};
\draw (-6,1) node {\footnotesize $\alpha_{1}$};
\draw (0,0) -- (6,0); 
\draw (2.5,1) node {\footnotesize $l_{1}$};
\draw (6,-5) -- (6,-2);
\fill[black] (6,-3.5) circle (0.25);
\fill[blue!80!red!20!] (6,0) circle (2);
\draw (6,0) circle (2) node {${\bf r}$};
\fill[blue!80!red!20!] (6,-7) circle (2);
\draw (6,-7) circle (2) node {${\bf p}$};
\end{move}
\end{DiagramV}
\end{align}
While this might look like some heavy-handed machinery, the utility of fiducial measurements and states (and thus, the first-level compression) will become evident when we start considering more regions.

\subsubsection{Tomographic Compression and Generalised Probability Theory}

The first level compression is deeply related to \emph{Generalised Probability Theories (GPTs)}. In the GPT framework, one can characterise physical theories by finding the relation between the number of distinguishable states for a system - N and the number of measurement outcomes required - K, to fully characterise the states of the system (when we do not include the normalisation condition).  In GPTs with a certain tomographic locality property, we have $K(N)=N^r$ where $r$ is theory dependent \cite{hardy2001quantum}. For finite-dimensional Classical Probability Theory $r=1$ while for finite-dimensional Quantum Theory $r=2$. If the underlying theory qualifies as a GPT the first level compression may be thought of as channel tomography and then the size of the fiducial measurement set is related to $K(N)$ as follows (where the channel input is denoted by $K_i(N_i)$, and channel output is denoted by $K_o(N_o)$):

\begin{align}
|\Omega_{R_1}| = K_iK_o = N_i^{r}N_o^{r} {\rm ~~where}  \left\{
  \begin{array}{lr}
   r=1 \quad\text{for Classical Probability Theory} \\
   r=2 \quad\text{for Quantum Theory} 
  \end{array}
\right.
\end{align}
Therefore, we will call the first level compression as \emph{Tomographic Compression}. We will also synonymously call the Omega sets, $\Omega_{R_1}$. for a region as Tomographic Sets and the Lambda matrices, $\Lambda$, for a region  as Tomographic Matrices. We will see more of Omega sets and Lambda matrices in the coming sections and they play a central role to the framework.

Note that we have focused on linear relations. One may ask why non-linear relations for physical compression are not considered.  Certainly, in principle, nonlinear compression might do better.  However, if we are able to form arbitrary mixtures of states like we can in Quantum Theory, then it is easy to prove that linear compression is optimal.  The reason for this is that probabilities combine in a linear fashion when we take such mixtures.  

\subsection{\label{compositionalcompression}Second level: Compositional Compression}

We have worked with Tomographic Compression that pertains to a single region, and the natural next question to ask is how do physical theories correlate data between multiple disjoint regions. For purposes of exposition let us explain using two regions.  Extending the following to multiple regions will be straightforward (discussed at the end of this subsection). Let us consider a composite region consisting of two disjoint regions $R_1$ and $R_2$ such that $R_1 \cap R_2 = \phi$ and $R_1, R_2 \subset R$ where $R$ is a larger predictively well-defined region. Since we are considering two regions the generalised preparations of states of one region will depend on the labels of the other region. Let $\alpha_1$ be the label for measurement outcomes in $R_1$ and $\alpha_2$ be the label for measurement outcomes in $R_2$. Then for $R_1$ we have
\begin{align}
     (Y^{\alpha_2}_{R_2},Y_{R-R_1-R_2}, F^{\alpha_2}_{R_2}, F_{R-R_1-R_2}) &\Longleftrightarrow \textit{generalised preparation for $R_1$}\\
 (Y^{\alpha_1}_{R_1}, F^{\alpha_1}_{R_1}) &\Longleftrightarrow \textit{measurement in $R_1$}
\end{align}
and similarly for $R_2$ we have
\begin{align}
     (Y^{\alpha_1}_{R_1},Y_{R-R_1-R_2}, F^{\alpha_1}_{R_1}, F_{R-R_1-R_2}) &\Longleftrightarrow \textit{generalised preparation for $R_2$}\\
 (Y^{\alpha_2}_{R_2}, F^{\alpha_2}_{R_2}) &\Longleftrightarrow \textit{measurement in $R_2$}
\end{align} 
For the composite region $R_1 \cup R_2$, the joint probabilities of interest are given by:
\begin{equation}\label{eq:pr1r2} p_{\alpha_1\alpha_2} = {\rm Prob}(Y^{\alpha_1}_{R_1}\cup
Y^{\alpha_2}_{R_2} \cup Y_{R-R_1-R_2}|F^{\alpha_1}_{R_1}\cup F^{\alpha_2}_{R_2} \cup F_{R-R_1-R_2})=
\begin{DiagramV}{3}{0}
\begin{move}{0,0}
\draw (0,4) -- (6,4); 
\draw (0,-4) -- (6,-4); 
\fill[blue!80!red!20!] (6,0) ellipse (2 and 6);
\draw (6,0) ellipse (2 and 6) node {$p$};
\draw (2.5,4+1) node {\footnotesize $\alpha_{1}$};
\draw (2.5,-4+1) node {\footnotesize $\alpha_{2}$};
\end{move}
\end{DiagramV}
\end{equation}
Given the only compression in our arsenal yet is the first, consider that we can either apply Tomographic compression to region $R_1$ and to region $R_2$, or apply Tomographic compression to the composite region $R_1 \cup R_2$. Will these two scenarios be equivalent? Let us find out.  

When we apply Tomographic compression to the region $R_1$ followed by Tomographic compression on $R_2$ we get:
\begin{eqnarray}
p_{\alpha_1\alpha_2} &=&{\bf r}_{\alpha_1}(R_1)\cdot {\bf p}_{\alpha_2}(R_1) \label{manipulationtworegions1}  \\
&=&\sum_{l_1\in\Omega_1} \Lambda_{\alpha_1}^{l_1} {\bf r}_{l_1}(R_1) \cdot {\bf p}_{\alpha_2}(R_1)  \label{manipulationtworegions2}  \\
&=&\sum_{l_1\in\Omega_1} \Lambda_{\alpha_1}^{l_1} { p}_{l_1\alpha_2}  \label{manipulationtworegions3}  
\end{eqnarray}
\begin{eqnarray}
&=&\sum_{l_1\in\Omega_1} \Lambda_{\alpha_1}^{l_1}{\bf r}_{\alpha_2}(R_2) \cdot {\bf p}_{l_1}(R_2)   \label{manipulationtworegions4}\\
&=&\sum_{l_1\in\Omega_1}\sum_{l_2\in\Omega_2} \Lambda_{\alpha_1}^{l_1}\Lambda_{\alpha_2}^{l_2}{\bf r}_{l_2}(R_2) \cdot {\bf p}_{l_1}(R_2)  \label{manipulationtworegions5}  \\
&=&\sum_{l_1l_2\in\Omega_1\times\Omega_2} \Lambda_{\alpha_1}^{l_1}\Lambda_{\alpha_2}^{l_2}{\bf r}_{l_2}(R_2) \cdot {\bf p}_{l_1}(R_2) \label{manipulationtworegions6}\\
&=&\sum_{l_1l_2\in\Omega_1\times\Omega_2} \Lambda_{\alpha_1}^{l_1}\Lambda_{\alpha_2}^{l_2} p_{l_1l_2}\label{manipulationtworegions7}
\end{eqnarray} 
where we have made repeated use of Equation \eqref{firstlevellambdar} for tomographic compression (from Equation \eqref{manipulationtworegions1} to \eqref{manipulationtworegions2} and from Equation \eqref{manipulationtworegions4} to \eqref{manipulationtworegions5}). One must pay close attention to the regions in brackets for the ${\bf r}$ and ${\bf p}$ vectors! ${\bf p}_{\alpha_2}(R_1)$ is the state for region $R_1$ labelled by generalised preparation label $\alpha_2$ in region $R_2$ and similarly ${\bf p}_{l_1}(R_2)$ is the state for region $R_2$ labelled by fiducial preparation label $l_1$ in region $R_1$, post Tomographic Compression. Finally, notice that we use:
\begin{align}
    l_1l_2\in\Omega_1\times\Omega_2
\end{align}
here $\times$ denotes the Cartesian product between the elements of the two sets $\Omega_1$ and $\Omega_2$, for example, $\{1,4,9\}\times\{1,2\}=\{11,12,41,42,91,92\}$.

The diagrammatic form for Equations \eqref{manipulationtworegions1}-\eqref{manipulationtworegions7} are given by:  
\begin{align}
\begin{DiagramV}{3}{0}
\begin{move}{0,0}
\draw (0,4) -- (6,4); 
\draw (0,-4) -- (6,-4); 
\fill[blue!80!red!20!] (6,0) ellipse (2 and 6);
\draw (6,0) ellipse (2 and 6) node {$p$};
\draw (2.5,4+1) node {\footnotesize $\alpha_{1}$};
\draw (2.5,-4+1) node {\footnotesize $\alpha_{2}$};
\end{move}
\end{DiagramV}
=&~~~
\begin{DiagramV}{3}{0}
\begin{move}{0,0}
\draw (0,4) -- (6,4); 
\draw (0,-4) -- (6,-4); 
\draw (6,-4) -- (6,4);
\fill[black] (6,0) circle (0.25);
\fill[blue!80!red!20!] (6,+4) circle (2);
\draw (6,+4) circle (2) node {${\bf r}$};
\fill[blue!80!red!20!] (6,-4) circle (2);
\draw (6,-4) circle (2) node {${\bf p}$};
\draw (2.5,4+1) node {\footnotesize $\alpha_{1}$};
\draw (2.5,-4+1) node {\footnotesize $\alpha_{2}$};
\end{move}
\end{DiagramV}
=
\begin{DiagramV}{3}{0}
\begin{move}{0,0}
\draw (0,4) -- (6,4); 
\draw (0,-4) -- (6,-4); 
\draw (6,-4) -- (6,4);
\fill[black] (6,0) circle (0.25);
\fill[blue!80!red!20!] (6,+4) circle (2);
\draw (6,+4) circle (2) node {${\bf r}$};
\fill[blue!80!red!20!] (6,-4) circle (2);
\draw (6,-4) circle (2) node {${\bf p}$};
\draw (2.5,-4+1) node {\footnotesize $\alpha_{2}$};
\draw (-8,4) -- (0,4); 
\fill[blue!80!red!20!] (-4,-2+4) rectangle (0,2+4);
\draw (-4,-2+4) rectangle (0,2+4);
\draw (-2,0+4) node {$\Lambda$};
\draw (-6,1+4) node {\footnotesize $\alpha_{1}$};
\draw (0,4) -- (4,4); 
\draw (2.5,1+4) node {\footnotesize $l_{1}$};
\end{move}
\end{DiagramV}\\
=&
\begin{DiagramV}{3}{0}
\begin{move}{0,0}
\draw (0,4) -- (6,4); 
\draw (0,-4) -- (6,-4); 
\fill[blue!80!red!20!] (6,0) ellipse (2 and 6);
\draw (6,0) ellipse (2 and 6) node {$p$};
\draw (-8,-4) -- (0,-4);
\draw (-6,-4+1) node {\footnotesize $\alpha_{2}$};
\draw (-8,4) -- (0,4); 
\fill[blue!80!red!20!] (-4,-2+4) rectangle (0,2+4);
\draw (-4,-2+4) rectangle (0,2+4);
\draw (-2,0+4) node {$\Lambda$};
\draw (-6,1+4) node {\footnotesize $\alpha_{1}$};
\draw (0,4) -- (4,4); 
\draw (2.5,1+4) node {\footnotesize $l_{1}$};
\end{move}
\end{DiagramV}
=
\begin{DiagramV}{3}{0}
\begin{move}{0,0}
\draw (0,4) -- (6,4); 
\draw (0,-4) -- (6,-4); 
 \draw (6,-4) -- (6,4);
 \fill[black] (6,0) circle (0.25);
\fill[blue!80!red!20!] (6,+4) circle (2);
\draw (6,+4) circle (2) node {${\bf p}$};
\fill[blue!80!red!20!] (6,-4) circle (2);
\draw (6,-4) circle (2) node {${\bf r}$};
\draw (-8,-4) -- (0,-4);
\draw (-6,-4+1) node {\footnotesize $\alpha_{2}$};
\draw (-8,4) -- (0,4); 
\fill[blue!80!red!20!] (-4,-2+4) rectangle (0,2+4);
\draw (-4,-2+4) rectangle (0,2+4);
\draw (-2,0+4) node {$\Lambda$};
\draw (-6,1+4) node {\footnotesize $\alpha_{1}$};
\draw (0,4) -- (4,4); 
\draw (2.5,1+4) node {\footnotesize $l_{1}$};
\end{move}
\end{DiagramV}\\
=&
\begin{DiagramV}{3}{0}
\begin{move}{0,0}
\draw (0,4) -- (6,4); 
\draw (0,-4) -- (6,-4); 
\draw (6,-4) -- (6,4);
\fill[black] (6,0) circle (0.25);
\fill[blue!80!red!20!] (6,+4) circle (2);
\draw (6,+4) circle (2) node {${\bf p}$};
\fill[blue!80!red!20!] (6,-4) circle (2);
\draw (6,-4) circle (2) node {${\bf r}$};
%
\draw (-8,4) -- (0,4); 
\fill[blue!80!red!20!] (-4,-2+4) rectangle (0,2+4);
\draw (-4,-2+4) rectangle (0,2+4);
\draw (-2,0+4) node {$\Lambda$};
\draw (-6,1+4) node {\footnotesize $\alpha_{1}$};
\draw (-8,-4) -- (0,-4); 
\fill[blue!80!red!20!] (-4,-2-4) rectangle (0,2-4);
\draw (-4,-2-4) rectangle (0,2-4);
\draw (-2,0-4) node {$\Lambda$};
\draw (-6,1-4) node {\footnotesize $\alpha_{2}$};
\draw (0,4) -- (4,4); 
\draw (2.5,1+4) node {\footnotesize $l_{1}$};
\draw (2.5,-4+1) node {\footnotesize $l_{2}$};
\end{move}
\end{DiagramV}
=
\begin{DiagramV}{3}{0}
\begin{move}{0,0}
\draw (0,4) -- (6,4); 
\draw (0,-4) -- (6,-4); 
\fill[black] (6,0) circle (0.25);
\fill[blue!80!red!20!] (6,0) ellipse (2 and 6);
\draw (6,0) ellipse (2 and 6) node {$p$};
\fill[blue!80!red!20!] (6,0) ellipse (2 and 6);
\draw (6,0) ellipse (2 and 6) node {$p$};
%
\draw (-8,4) -- (0,4); 
\fill[blue!80!red!20!] (-4,-2+4) rectangle (0,2+4);
\draw (-4,-2+4) rectangle (0,2+4);
\draw (-2,0+4) node {$\Lambda$};
\draw (-6,1+4) node {\footnotesize $\alpha_{1}$};
\draw (-8,-4) -- (0,-4); 
\fill[blue!80!red!20!] (-4,-2-4) rectangle (0,2-4);
\draw (-4,-2-4) rectangle (0,2-4);
\draw (-2,0-4) node {$\Lambda$};
\draw (-6,1-4) node {\footnotesize $\alpha_{2}$};
\draw (0,4) -- (4,4); 
\draw (2.5,1+4) node {\footnotesize $l_{1}$};
\draw (2.5,-4+1) node {\footnotesize $l_{2}$};
\end{move}
\end{DiagramV}
\end{align}

If instead one chooses to tomographically compress $R_2$ first and $R_1$ later then the analogous calculation to Equations \eqref{manipulationtworegions1}-\eqref{manipulationtworegions7} will include the following step in place of \eqref{manipulationtworegions6}:
\begin{eqnarray}\label{manipulationtworegionsorderswapped}
p_{\alpha_1\alpha_2} =\sum_{l_1l_2\in\Omega_1\times\Omega_2} \Lambda_{\alpha_1}^{l_1}\Lambda_{\alpha_2}^{l_2}{\bf r}_{l_1}(R_1) \cdot {\bf p}_{l_2}(R_1)
\end{eqnarray}
but the order of first-level compression to disjoint regions does not matter and comparing Equation \eqref{manipulationtworegions6} and Equation \eqref{manipulationtworegionsorderswapped} we have:
\begin{eqnarray}\label{manipulationtworegionsprobability}
p_{l_1l_2} ={\bf r}_{l_1}(R_1) \cdot {\bf p}_{l_2}(R_1)={\bf r}_{l_2}(R_2) \cdot {\bf p}_{l_1}(R_2) \nonumber \\
\text{such that} \quad p_{\alpha_1\alpha_2} =\sum_{l_1l_2\in\Omega_1\times\Omega_2} \Lambda_{\alpha_1}^{l_1}\Lambda_{\alpha_2}^{l_2}p_{l_1l_2}
\end{eqnarray}

Let us now consider the second option, directly applying tomographic compression to the composite region
\begin{align}
    R_{1,2}=R_1 \cup R_2
\end{align}
The state for $R_{1,2}$ is any mathematical object that can be used to calculate all $p_{\alpha_1\alpha_2}$ by means of linear relations. The state for $R_{1,2}$ after performing Tomographic Compression on the composite region is defined as:
\begin{equation}
{\bf p}(R_{1,2}) = \left( \begin{array}{c} \vdots \\ p_{k_1k_2}\\ \vdots \end{array} \right) ~~\text{, where}~~k_1k_2\in\Omega_{1,2}
\end{equation}
where $\Omega_{1,2}$ is tomographic set for $R_{1,2}$. For some choice, the fiducial measurements are given by:
\begin{align}
    {\bf r}_{k_1k_2}= \left( \begin{array}{c} 0 \\\vdots \\ 1 \\ \vdots \\0 \end{array}  \right)\quad\text{for all}\quad k_1k_2 \in \Omega_{1,2}
\end{align}
where, the $k_1k_2^{th}$ element is $1$ and the rest are $0$. The state and fiducial measurements can then be used to calculate probabilities using the generalised Born rule:
\begin{align}\label{raadotpaa}
p_{\alpha_1\alpha_2} &= {\bf r}_{\alpha_1\alpha_2}(R_{1,2})\cdot {\bf p}(R_{1,2})\\
                     &= \sum_{k_1k_2\in \Omega_{1,2}} \Lambda^{k_1k_2}_{\alpha_1\alpha_2}{\bf r}_{k_1k_2}(R_{1,2})\cdot {\bf p}(R_{1,2})
\end{align}
Equivalently and diagrammatically we have
\begin{align}
\begin{DiagramV}{3}{0}
\begin{move}{0,0}
\draw (0,4) -- (6,4); 
\draw (0,-4) -- (6,-4); 
\fill[blue!80!red!20!] (6,0) ellipse (2 and 5);
\draw (6,0) ellipse (2 and 5) node {$p$};
\draw (2.5,4+1) node {\footnotesize $\alpha_{1}$};
\draw (2.5,-4+1) node {\footnotesize $\alpha_{2}$};
\end{move}
\end{DiagramV}
=
\begin{DiagramV}{3}{0}
\begin{move}{0,0}
\draw (0,4) -- (6,4); 
\draw (0,-4) -- (6,-4); 
\draw (6,-9) -- (6,-2);
\fill[blue!80!red!20!] (6,0) ellipse (2 and 5);
\draw (6,0) ellipse (2 and 5) node {${\bf r}$};
\draw (2.5,4+1) node {\footnotesize $\alpha_{1}$};
\draw (2.5,-4+1) node {\footnotesize $\alpha_{2}$};
\fill[black] (6,-6) circle (0.25);
\fill[blue!80!red!20!] (6,-9) circle (2);
\draw (6,-9) circle (2) node {${\bf p}$};
\end{move}
\end{DiagramV}
=
\begin{DiagramV}{3}{0}
\begin{move}{0,0}
\draw (0,4) to[out=0,in=180] (4,1); 
\draw (0,-4) to[out=0,in=180] (4,-1); 
\draw (6,0) -- (6,-10); 
 \draw (-8,-4) -- (0,-4); 
\fill[black] (6,0) circle (0.25);
\fill[blue!80!red!20!] (6,0) ellipse (2 and 3);
\draw (6,0) ellipse (2 and 3) node {$p$};
\fill[blue!80!red!20!] (6,0) ellipse (2 and 3);
\draw (6,0) ellipse (2 and 3) node {${\bf r}$};
\fill[black] (6,-6) circle (0.25);
\fill[blue!80!red!20!] (6,-9) circle (2);
\draw (6,-9) circle (2) node {${\bf p}$};
%
\draw (-8,4) -- (0,4); 
\fill[blue!80!red!20!] (-4,-2-4) rectangle (0,2+4);
\draw (-4,-2-4) rectangle (0,2+4);
\draw (-2,0) node {$\Lambda$};
\draw (-6,1+4) node {\footnotesize $\alpha_{1}$};
\draw (-6,1-4) node {\footnotesize $\alpha_{2}$};
\draw (2.5,4) node {\footnotesize $k_{1}$};
\draw (2.5,-4) node {\footnotesize $k_{2}$};
\end{move}
\end{DiagramV}
\end{align}

We are ready to compare how the compression is quantified in both cases, by comparing the $\Omega$ sets and $\Lambda$ matrices. We have $\Lambda^{l_1}_{\alpha_1}\Lambda^{l_2}_{\alpha_2}$ where $l_1l_2 \in \Omega_1 \times \Omega_2$ when tomographically compressing each region separately. And we have $\Lambda^{k_1k_2}_{\alpha_1\alpha_2}$ where $k_1k_2 \in \Omega_{1,2}$ when tomographically compressing the composite region.

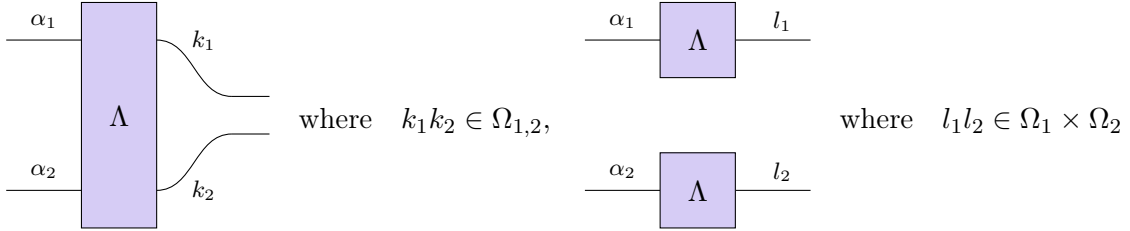
\begin{align}
    \begin{DiagramV}{0}{0}
    \begin{move}{0,0}
    \draw (0,4) to[out=0,in=180] (4,1); 
    \draw (0,-4) to[out=0,in=180] (4,-1); 
    \draw (4,-1) -- (6,-1); 
    \draw (4,1) -- (6,1); 
    \draw (-8,4) -- (0,4);
    \draw (-8,-4) -- (0,-4);
    \fill[blue!80!red!20!] (-4,-2-4) rectangle (0,2+4);
    \draw (-4,-2-4) rectangle (0,2+4);
    \draw (-2,0) node {$\Lambda$};
    \draw (-6,1+4) node {\footnotesize $\alpha_{1}$};
    \draw (-6,1-4) node {\footnotesize $\alpha_{2}$};
    \draw (2.5,4) node {\footnotesize $k_{1}$};
    \draw (2.5,-4) node {\footnotesize $k_{2}$};
    \end{move}
    \end{DiagramV}
\quad \text{where} \quad k_1k_2\in\Omega_{1,2}, \quad
    \begin{DiagramV}{0}{0}
    \begin{move}{0,0}
    \draw (0,4) -- (4,4); 
    \draw (0,-4) -- (4,-4); 
    \draw (-8,4) -- (0,4); 
    \fill[blue!80!red!20!] (-4,-2+4) rectangle (0,2+4);
    \draw (-4,-2+4) rectangle (0,2+4);
    \draw (-2,0+4) node {$\Lambda$};
    \draw (-6,1+4) node {\footnotesize $\alpha_{1}$};
    \draw (-8,-4) -- (0,-4); 
    \fill[blue!80!red!20!] (-4,-2-4) rectangle (0,2-4);
    \draw (-4,-2-4) rectangle (0,2-4);
    \draw (-2,0-4) node {$\Lambda$};
    \draw (-6,1-4) node {\footnotesize $\alpha_{2}$};
    \draw (2.5,1+4) node {\footnotesize $l_{1}$};
    \draw (2.5,-4+1) node {\footnotesize $l_{2}$};
    \end{move}
    \end{DiagramV} 
\quad \text{where} \quad l_1l_2\in\Omega_{1}\times\Omega_2\nonumber
\end{align}
It is evident that any compression achieved by $\Lambda^{l_1}_{\alpha_1}\Lambda^{l_2}_{\alpha_2}$ will at least be achieved by $\Lambda^{k_1k_2}_{\alpha_1\alpha_2}$, since it can take the form $\Lambda^{k_1k_2}_{\alpha_1\alpha_2}=\Lambda^{k_1}_{\alpha_1}\Lambda^{k_2}_{\alpha_2}$. In fact, $\Lambda^{k_1k_2}_{\alpha_1\alpha_2}$ will sometimes provide further compression that cannot be found by tomographic compression of the region $R_1$ and tomographic compression of the region $R_2$. Therefore we have the following result which is central to this framework:
\begin{equation}\label{centralresult}
\Omega_{1,2} \subseteq \Omega_1\times\Omega_2
\end{equation}

We may finally define the second level or Compositional Compression for composite regions as the compression that is found over and above the first level or Tomographic Compression of constituent regions. We define the Compositional Lambda matrix as follows:
\begin{align}
    {\bf r}_{l_1l_2}&=\sum_{k_1,k_2\in\Omega_{1,2}}\Lambda^{k_1k_2}_{l_1l_2}{\bf r}_{k_1k_2}\label{eqn:secondlevel} \\
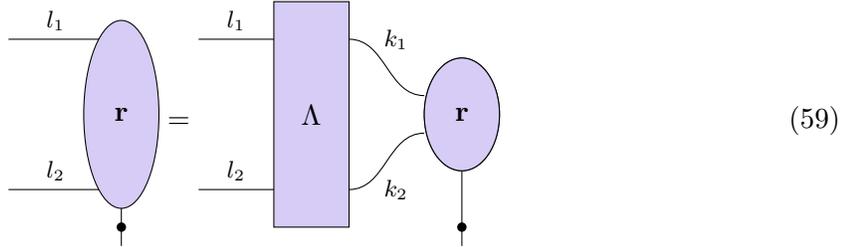
    \begin{DiagramV}{2}{0}
    \begin{move}{0,0}
    \draw (0,4) -- (6,4); 
    \draw (0,-4) -- (6,-4); 
    \draw (6,-7) -- (6,-2);
    \fill[blue!80!red!20!] (6,0) ellipse (2 and 5);
    \draw (6,0) ellipse (2 and 5) node {${\bf r}$};
    \draw (2.5,4+1) node {\footnotesize $l_1$};
    \draw (2.5,-4+1) node {\footnotesize $l_2$};
    \fill[black] (6,-6) circle (0.25);
    \end{move}
    \end{DiagramV}
    &=
    \begin{DiagramV}{0}{0}
    \begin{move}{0,0}
    \draw (0,4) to[out=0,in=180] (4,1); 
    \draw (0,-4) to[out=0,in=180] (4,-1); 
    \draw (6,0) -- (6,-7); 
     \draw (-8,-4) -- (0,-4); 
    \fill[black] (6,0) circle (0.25);
    \fill[blue!80!red!20!] (6,0) ellipse (2 and 3);
    \draw (6,0) ellipse (2 and 3) node {$p$};
    \fill[blue!80!red!20!] (6,0) ellipse (2 and 3);
    \draw (6,0) ellipse (2 and 3) node {${\bf r}$};
    \fill[black] (6,-6) circle (0.25);
    \draw (-8,4) -- (0,4); 
    \fill[blue!80!red!20!] (-4,-2-4) rectangle (0,2+4);
    \draw (-4,-2-4) rectangle (0,2+4);
    \draw (-2,0) node {$\Lambda$};
    \draw (-6,1+4) node {\footnotesize $l_{1}$};
    \draw (-6,1-4) node {\footnotesize $l_{2}$};
    \draw (2.5,4) node {\footnotesize $k_{1}$};
    \draw (2.5,-4) node {\footnotesize $k_{2}$};
    \end{move}
    \end{DiagramV}
\end{align}
where $\Lambda_{l_1l_2}^{k_1k_2}$ encodes Compositional Compression. Here, $l_1,l_2 \in \Omega_1\times\Omega_2$ are the labels after tomographic compression of each constituent region, conventionally we will use $l$'s for labelling tomographic compression. Compositional Compression is nontrivial when $\Omega_{1,2}$ is a proper subset of $\Omega_1\times\Omega_2$ and conventionally we will use $k$'s for compositional compression.  

Compositional Compression for multiple disjoint regions can be implemented through an extension of the examples discussed above. Multi-region compression will go through the following label changes
\begin{equation}
l_1l_2\dots l_n \in \Omega_1\times\Omega_2\times\dots\times\Omega_n  \longrightarrow k_1k_2\dots k_n \in \Omega_{1,2\dots n} 
\end{equation}
and the associated Lambda matrix, which encodes Compositional Compression takes the form 
\begin{equation}
\Lambda_{l_1l_2\dots l_n}^{k_1k_2\dots k_n}
\end{equation}

This concludes the details for the second-level compression. Let us recapitulate the steps in applying Compositional Compression. The person in the box is interested in $n$-regions, each associated with the label $\alpha_i$ where $i$ specifies the region. Therefore the person is interested in probabilities of the form $p_{\alpha_1...\alpha_n}$. They perform tomographic compression on each region. Then they perform compositional compression on the composite region. Mathematically:

\begin{align}
p_{\alpha_1\alpha_2 \dots \alpha_n} &= \mathbf{r}_{\alpha_1 \alpha_2\dots \alpha_n} \cdot \mathbf{p} \\
&= \sum_{l_1l_2\dots l_n \in \Omega_1 \times \Omega_2 \times \dots \times\Omega_n} \Lambda_{\alpha_1}^{l_1}\Lambda_{\alpha_2}^{l_2}\dots\Lambda_{\alpha_n}^{l_n}\mathbf{r}_{l_1l_2\dots l_n} \cdot \mathbf{p} \\
&= \sum_{l_1l_2\dots l_n \in \Omega_1 \times \Omega_2 \times \dots \times\Omega_n} \Lambda_{\alpha_1}^{l_1}\Lambda_{\alpha_2}^{l_2}\dots\Lambda_{\alpha_n}^{l_n} \sum_{k_1k_2 \dots k_n \in \Omega_{1,2,\dots,n}} \Lambda_{l_1l_2 \dots l_n}^{k_1k_2 \dots k_n} \mathbf{r}_{k_1k_2\dots k_n} \cdot \mathbf{p}
\end{align}
and diagrammatically:
\begin{align}
\begin{DiagramV}{3}{0}
\begin{move}{0,0}
\draw (0,4) -- (6,4); 
\draw (0,1) -- (6,1); 
\draw (0,-4) -- (6,-4); 
\fill[blue!80!red!20!] (6,0) ellipse (2 and 5);
\draw (6,0) ellipse (2 and 5) node {$p$};
\draw (2.5,4+1) node {\footnotesize $\alpha_{1}$};
\draw (2.5,1+1) node {\footnotesize $\alpha_{2}$};
\draw (2.5,-1.5+1) node {\footnotesize $\vdots$};
\draw (2.5,-4+1) node {\footnotesize $\alpha_{n}$};
\end{move}
\end{DiagramV}
&=
\begin{DiagramV}{3}{0}
\begin{move}{0,0}
\draw (0,4) -- (6,4); 
\draw (0,1) -- (6,1); 
\draw (0,-4) -- (6,-4); 
\draw (6,-9) -- (6,-2);
\fill[blue!80!red!20!] (6,0) ellipse (2 and 5);
\draw (6,0) ellipse (2 and 5) node {${\bf r}$};
\draw (2.5,4+1) node {\footnotesize $\alpha_{1}$};
\draw (2.5,1+1) node {\footnotesize $\alpha_{2}$};
\draw (2.5,-1.5+1) node {\footnotesize $\vdots$};
\draw (2.5,-4+1) node {\footnotesize $\alpha_{n}$};
\fill[black] (6,-6) circle (0.25);
\fill[blue!80!red!20!] (6,-9) circle (2);
\draw (6,-9) circle (2) node {${\bf p}$};
\end{move}
\end{DiagramV}
=
\begin{DiagramV}{3}{0}
\begin{move}{0,0}
\draw (-6,4) -- (6,4); 
\draw (-6,1) -- (6,1); 
\draw (-6,-4) -- (6,-4); 
\draw (6,-9) -- (6,-2);
\fill[blue!80!red!20!] (6,0) ellipse (2 and 5);
\draw (6,0) ellipse (2 and 5) node {${\bf r}$};
\draw (2.5,4+1) node {\footnotesize $l_{1}$};
\draw (2.5,1+1) node {\footnotesize $l_{2}$};
\draw (2.5,-1.5+1) node {\footnotesize $\vdots$};
\draw (2.5,-4+1) node {\footnotesize $l_{n}$};
\draw (-4,4+1) node {\footnotesize $\alpha_{1}$};
\draw (-4,1+1) node {\footnotesize $\alpha_{2}$};
\draw (-4,-1.5+1) node {\footnotesize $\vdots$};
\draw (-4,-4+1) node {\footnotesize $\alpha_{n}$};
\fill[black] (6,-6) circle (0.25);
\fill[blue!80!red!20!] (6,-9) circle (2);
\draw (6,-9) circle (2) node {${\bf p}$};
\fill[blue!80!red!20!] (-2,-1+4) rectangle (0,1+4);
\draw (-2,-1+4) rectangle (0,1+4);
\draw (-1,+4) node {\footnotesize $\Lambda$};
\fill[blue!80!red!20!] (-2,-1+1) rectangle (0,1+1);
\draw (-2,-1+1) rectangle (0,1+1);
\draw (-1,+1) node {\footnotesize $\Lambda$};
\fill[blue!80!red!20!] (-2,-1-4) rectangle (0,1-4);
\draw (-2,-1-4) rectangle (0,1-4);
\draw (-1,-4) node {\footnotesize $\Lambda$};
\end{move}
\end{DiagramV} \\
&=
\begin{DiagramV}{3}{0}
\begin{scope}[shift={(-8,0)}]
\draw (-6,4) -- (6,4); 
\draw (-6,1) -- (6,1); 
\draw (-6,-4) -- (6,-4); 
\draw (2.5,4+1) node {\footnotesize $l_{1}$};
\draw (2.5,1+1) node {\footnotesize $l_{2}$};
\draw (2.5,-1.5+1) node {\footnotesize $\vdots$};
\draw (2.5,-4+1) node {\footnotesize $l_{n}$};
\draw (-4,4+1) node {\footnotesize $\alpha_{1}$};
\draw (-4,1+1) node {\footnotesize $\alpha_{2}$};
\draw (-4,-1.5+1) node {\footnotesize $\vdots$};
\draw (-4,-4+1) node {\footnotesize $\alpha_{n}$};
\fill[blue!80!red!20!] (-2,-1+4) rectangle (0,1+4);
\draw (-2,-1+4) rectangle (0,1+4);
\draw (-1,+4) node {\footnotesize $\Lambda$};
\fill[blue!80!red!20!] (-2,-1+1) rectangle (0,1+1);
\draw (-2,-1+1) rectangle (0,1+1);
\draw (-1,+1) node {\footnotesize $\Lambda$};
\fill[blue!80!red!20!] (-2,-1-4) rectangle (0,1-4);
\draw (-2,-1-4) rectangle (0,1-4);
\draw (-1,-4) node {\footnotesize $\Lambda$};
\end{scope}
\begin{move}{0,0}
\draw (0,4) to[out=0,in=180] (4,1); 
\draw (0,1) to[out=0,in=180] (4,0.5); 
\draw (0,-4) to[out=0,in=180] (4,-1); 
\draw (6,0) -- (6,-10); 
\fill[black] (6,0) circle (0.25);
\fill[blue!80!red!20!] (6,0) ellipse (2 and 3);
\draw (6,0) ellipse (2 and 3) node {${\bf r}$};
\fill[black] (6,-6) circle (0.25);
\fill[blue!80!red!20!] (6,-9) circle (2);
\draw (6,-9) circle (2) node {${\bf p}$};
\fill[blue!80!red!20!] (-4,-1-4) rectangle (0,1+4);
\draw (-4,-1-4) rectangle (0,1+4);
\draw (-2,0) node {$\Lambda$};
\draw (1,5) node {\footnotesize $k_{1}$};
\draw (1,2) node {\footnotesize $k_{2}$};
\draw (1,-0.5) node {\footnotesize $\vdots$};
\draw (1,-3+0.7) node {\footnotesize $k_{n}$};
\end{move}
\end{DiagramV}
\end{align}

Similar to ${\bf r}_{\alpha_1}|_{l_1}$, we can show that the components of the Lambda matrices can be specified through entries of the ${\bf r}$-vectors. 
\begin{align}\label{rsecondlevelcomponent}
    {\bf r}_{l_1l_2\dots l_n}|_{k_1k_2\dots k_n} :=& \Lambda_{l_1l_2 \dots l_n}^{k_1k_2 \dots k_n}=
       \begin{DiagramV}{3}{0}
    \begin{scope}[shift={(-8,0)}]
    \draw (0,4) -- (6,4); 
    \draw (0,1) -- (6,1); 
    \draw (0,-4) -- (6,-4); 
    \draw (2.5,4+1) node {\footnotesize $l_{1}$};
    \draw (2.5,1+1) node {\footnotesize $l_{2}$};
    \draw (2.5,-1.5+1) node {\footnotesize $\vdots$};
    \draw (2.5,-4+1) node {\footnotesize $l_{n}$};
    \end{scope}
    \begin{move}{0,0}
    \draw (0,4) to[out=0,in=180] (4,1); 
    \draw (0,1) to[out=0,in=180] (4,0.5); 
    \draw (0,-4) to[out=0,in=180] (4,-1); 
    \fill[blue!80!red!20!] (-4,-1-4) rectangle (0,1+4);
    \draw (-4,-1-4) rectangle (0,1+4);
    \draw (-2,0) node {$\Lambda$};
    \draw (1,5) node {\footnotesize $k_{1}$};
    \draw (1,2) node {\footnotesize $k_{2}$};
    \draw (1,-0.5) node {\footnotesize $\vdots$};
    \draw (1,-3+0.7) node {\footnotesize $k_{n}$};
    \end{move}
    \end{DiagramV}
    \\
    {\bf r}_{\alpha_1\alpha_2\dots \alpha_n}|_{k_1k_2\dots k_n}:=&\sum_{\substack{l_1l_2\dots l_n \\\in \Omega_1 \times \Omega_2 \times \dots \times\Omega_n}} \Lambda_{\alpha_1}^{l_1}\Lambda_{\alpha_2}^{l_2}\dots\Lambda_{\alpha_n}^{l_n} \Lambda_{l_1l_2 \dots l_n}^{k_1k_2 \dots k_n}
    =
    \begin{DiagramV}{3}{0}
    \begin{scope}[shift={(-8,0)}]
    \draw (-6,4) -- (6,4); 
    \draw (-6,1) -- (6,1); 
    \draw (-6,-4) -- (6,-4); 
    \draw (2.5,4+1) node {\footnotesize $l_{1}$};
    \draw (2.5,1+1) node {\footnotesize $l_{2}$};
    \draw (2.5,-1.5+1) node {\footnotesize $\vdots$};
    \draw (2.5,-4+1) node {\footnotesize $l_{n}$};
    \draw (-4,4+1) node {\footnotesize $\alpha_{1}$};
    \draw (-4,1+1) node {\footnotesize $\alpha_{2}$};
    \draw (-4,-1.5+1) node {\footnotesize $\vdots$};
    \draw (-4,-4+1) node {\footnotesize $\alpha_{n}$};
    \fill[blue!80!red!20!] (-2,-1+4) rectangle (0,1+4);
    \draw (-2,-1+4) rectangle (0,1+4);
    \draw (-1,+4) node {\footnotesize $\Lambda$};
    \fill[blue!80!red!20!] (-2,-1+1) rectangle (0,1+1);
    \draw (-2,-1+1) rectangle (0,1+1);
    \draw (-1,+1) node {\footnotesize $\Lambda$};
    \fill[blue!80!red!20!] (-2,-1-4) rectangle (0,1-4);
    \draw (-2,-1-4) rectangle (0,1-4);
    \draw (-1,-4) node {\footnotesize $\Lambda$};
    \end{scope}
    \begin{move}{0,0}
    \draw (0,4) to[out=0,in=180] (4,1); 
    \draw (0,1) to[out=0,in=180] (4,0.5); 
    \draw (0,-4) to[out=0,in=180] (4,-1); 
    \fill[blue!80!red!20!] (-4,-1-4) rectangle (0,1+4);
    \draw (-4,-1-4) rectangle (0,1+4);
    \draw (-2,0) node {$\Lambda$};
    \draw (1,5) node {\footnotesize $k_{1}$};
    \draw (1,2) node {\footnotesize $k_{2}$};
    \draw (1,-0.5) node {\footnotesize $\vdots$};
    \draw (1,-3+0.7) node {\footnotesize $k_{n}$};
    \end{move}
    \end{DiagramV}
\end{align}


\subsubsection{Causal Adjacency and $\Omega_{1,2}$}

Let us take a moment to discuss the implications of Equation \eqref{centralresult}. In what physical situations do we expect to (not) see nontrivial Compositional Compression? Suppose two regions are spatially separated and causally disconnected. In that case, we cannot reduce the number of parameters required to describe them, such as the tensor product in quantum theory, or spatially separated events in relativity - in that case, we have $\Omega_{1,2}=\Omega_{1}\times \Omega_2$. Even if a causal connection is possible but is mediated through other regions then these other regions can break the correlations between the two regions of interest. Consider for example, three quantum gates applied sequentially (like the polariser example from before), then the correlations between the first and third gate will depend on the nature of the second gate (if the second polariser is replaced by a cardboard and a new light is sent out then correlations between the first and third polarisers will be broken). In this case as well we have $\Omega_{1,2}=\Omega_{1}\times \Omega_2$. 

On the other hand, if we have strong causal connections between two regions that are not contingent on other regions (for example two consecutive quantum gates, which we can replace with a new composite quantum gate), then we will see that $\Omega_{1,2}$ will be a proper subset of $\Omega_{1}\times \Omega_2$. We will then say that the two regions are \emph{causally adjacent} (which we will denote by $R_1 \bowtie R_2)$. 

When many regions are of interest we can use Compositional Compression between different composite regions to map out causal adjacency. Therefore, Compositional Compression provides us with the ``mathematical signature for causal structure\rq\rq. We will see some more involved examples of Compositional Compression and a visualisation tool for Omega sets in the follow-up work \cite{sakharwade2023hierarchy} where we set up the hierarchy and consider multiple regions, and again later in the follow-up work \cite{sakharwade2023duotensors} we see some explicit examples about the Duotensor Framework (and in turn finite dimensional Quantum Theory). 


\subsection{\label{metacompression}Third level: Meta Compression} 

The previous two levels of compression pertain to the compression of single and multiple regions (which lie within a larger predictively well-defined region $R$), that are encoded combinatorially through the Omega sets and quantitatively through the Lambda matrices. But are the Lambda matrices for different regions independent of each other? A physical theory will in fact often have some structure that will correlate the Lambda matrices itself. The third level of compression will capture this compression of the Lambda matrices, through what is called \emph{the Causaloid}, one of the central mathematical objects in the Causaloid approach, owing to the framework its name. It is defined as:
\begin{quote}
{\bf The Causaloid } for a predictively well-defined region $R$ made up of  elementary regions $R_x$ is defined to
be that thing represented by any mathematical object which can be used to obtain ${\bf r}_{\alpha_{\cal O}}(R_{\cal
O})$ for all measurements $\alpha_{\cal O}$ in region $R_{\cal O}$ for all $R_{\cal O}\subseteq R$ (from \cite{hardy2005probability}). 
\end{quote}
To begin, let us find a mathematical object that specifies the Causaloid for a predictively well-defined region $R$, utilising the two levels of compression. The smallest component regions of $R$ are the elementary regions $R_x$, such that $\cup_{x} R_x =R$. $R$ can be expressed as consisting non-elementary regions $R_{\mathcal{O}}$ which itself consist of elementary regions. The region $R_{\mathcal{O}}$ can be specified in terms of elementary regions in the following way:
\begin{align}
    R_{\mathcal{O}}= \bigcup_{i\in\mathcal{O}} R_{i} \quad \text{where} \quad \mathcal{O}=\{x,x',...,x''\}
\end{align}
Of course $R_{\mathcal{O}}$ can reduce to a elementary region if $\mathcal{O}=\{x\}$. Now let us discuss the general region $R_{\mathcal{O}}$ through tomographic and compositional compression. In the pre-compression level the general measurement in the region $R_{\mathcal{O}}$ is labelled by  $\alpha_{\cal O}$, which decomposes into local measurements at each elementary region part of ${\cal O}$:
\begin{equation}
\alpha_{\cal O} = \alpha_{x}\alpha_{x'}\cdots \alpha_{x''}
\end{equation}
Each region $R_{\mathcal{O}}$ is associated with tomographic compression. Each composite region $R_{\mathcal{O}}$ (say for $R_{\mathcal{O}}=R_{\mathcal{O'}}\cup R_{\mathcal{O''}}\cup...$) is associated with compositional compression. But remember that we discussed that the tomographic compression of a composite region is equivalent to the tomographic compression of its constituents along with compositional compression on the composite region. In light of this, it is sufficient to consider tomographic compression of all elementary regions $R_x$ and compositional compression over all composite regions $R_{\mathcal{O}}$. The tomographic compression of an elementary region is given by: 
\begin{align}
    \Lambda_{\alpha_{x}}^{l_{x}}(x,\Omega_{x})
\end{align}
where the Lambda matrix is a function of $(x,\Omega_{x})$ -- the region and the Omega set given that it may not be unique. Similarly, the compositional compression over non-elementary regions $R_{\mathcal{O}}$ is given by
\begin{align}
    \Lambda_{l_{\cal O}}^{k_{\cal O}}(\mathcal{O},\Omega_{\cal O})
\end{align}
where, similar to the label $\alpha_{\cal O}$, the tomographic Omega set label $l_{\cal O}$ for the fiducial measurements of the composite region will also decompose into the fiducial measurements of the constituent elementary regions, while the compositional Omega set label $k_{\cal O}$ will not. That is:
\begin{align}
l_{\cal O} \equiv l_{x} l_{x'}\cdots l_{x''}\in\Omega_{x}\times\Omega_{x'}\times \cdots\times \Omega_{x''} \quad \text{and}\quad
k_{\cal O} \equiv k_{x} k_{x'}\cdots k_{x''}\in\Omega_{\cal O} 
\end{align}

Note that if $\mathcal{O}=\{x\}$, then there is no compositional compression since we are considering a single tomographically compressed region, and thus only a single Omega set $\Omega_x$ is relevant:
\begin{align}
    \Lambda_{l_{x}}^{k_{x}}=\delta_{l_{x}}^{k_{x}} \quad \text{since} \quad l_{x},k_{x}\in \Omega_x
\end{align}

It is worth noting that fully specifying the Lambda matrix $\Lambda_{l_{\cal O}}^{k_{\cal O}}({\cal O}, \Omega_{\cal O})$ would include specifying the region and a choice of Omega set, but it is easy to transform the Lambda matrix from one set of Omega choice to another using the following relation:
\begin{equation}
\Lambda_{l_{\cal O}}^{k'_{\cal O}}({\cal O}, \Omega'_{\cal O}) = \sum_{k_{\cal O}}[\Lambda_{k'_{\cal O}}^{k_{\cal
O}}({\cal O}, \Omega_{\cal O})]^{-1} \Lambda_{l_{\cal O}}^{k_{\cal O}}({\cal O}, \Omega_{\cal O})
\end{equation}
where
\begin{equation}\label{omegaomegaprime}
\Lambda_{k'_{\cal O}}^{k_{\cal O}}({\cal O}, \Omega_{\cal O}) ~~~~ k'_{\cal O} \in \Omega'_{\cal O}
\end{equation}
is a square matrix whose inverse exists, given two choices of Omega sets $\Omega_{\cal O}, \Omega'_{\cal O}$. Note that the matrix is square since $|\Omega_{\cal O}|=|\Omega'_{\cal O}|$. Also note that $k'_{\mathcal{O}}$ and $k_{\mathcal{O}}$ both belong to (different) subsets of $\Omega_{x}\times\Omega_{x'}\times\dots\times\Omega_{x''}$ since there is always a $l_{\mathcal{O}}$ for every $k'_{\mathcal{O}}$ and $k_{\mathcal{O}}$. A similar argument gives the relation for the transformation of tomographic Lambda matrices $\Lambda_{\alpha_x}^{l_x}(x,\Omega_x)$ of elementary regions. The transformation rule of Lambdas tells us that considering any one choice of Omega set for each region $\mathcal{O}$ is sufficient for the specification of the Causaloid.

Given these details we can write down one mathematical object that surely specifies the Causaloid (denoted by ${\bf \Lambda}$):
\begin{equation}\label{bigcausaloid}
{\bf \Lambda}=\left[
\begin{array}{ll} \Lambda_{\alpha_x}^{l_x}(x, \Omega_x) &: {\rm for~a} ~~\Omega_x ~~{\rm for~each~elementary} ~~ R_x \\
\Lambda^{k_{\cal O}}_{l_{\cal O}}({\cal O}, \Omega_{\cal O})&: {\rm for~a} ~~\Omega_{\cal O} ~\text{for each
non-elementary} ~ R_{\cal O} \subseteq R
\end{array}
\right]
\end{equation}
This satisfies the definition of the Causaloid since the lambda matrices can be used to calculate any ${\bf r}$-vector using the results stated in previous subsections. From Equations \eqref{rfirstlevelcomponent} and \eqref{rsecondlevelcomponent} we have \begin{equation}\label{rmultilambdax}
{\bf r}_{\alpha_x}|_{l_x} = \Lambda_{\alpha_x}^{l_x}  ~~~~\text{and}~~~~ {\bf r}_{\alpha_{\cal O}}|_{l_{\cal O}} =
\sum_{l_{\cal O}} \Lambda_{\alpha_x}^{l_x} \Lambda_{\alpha_{x'}}^{l_{x'}}\cdots \Lambda_{\alpha_{x''}}^{l_{x''}}
\Lambda_{l_{\cal O}}^{k_{\cal O}}
\end{equation}
This specification of the Causaloid is independent of any physical theory. But when considering a particular physical theory the specification of the Causaloid can be compressed given some structure. In such a scenario we expect to be able to calculate some $\Lambda$ matrices from others and thus, we can
take some subset of the $\Lambda$ matrices. Let us label such a subset by $i$, then we have
\begin{equation}\label{compressedCausaloid}
{\bf \Lambda}= [ \Lambda(i): i=1~{\rm to}~ I~|~\textit{rules}~]
\end{equation}
where \emph{rules} prescribe us the rules for deducing all $\Lambda$'s from the given subset of $\Lambda(i)$'s. The subset may not be unique, but it will tell us some key aspects of the physical theory.  Once we know the Causaloid we can calculate any probability within the physical theory.  Hence, the Causaloid does, in some sense, specify the physical theory itself.  It does so without imposing a definite causal structure.    

We may regard going from Equation \eqref{bigcausaloid} to \eqref{compressedCausaloid} as a kind of physical compression. We name this third level of compression as \emph{Meta Compression} since it does not look like the first two levels and acts on sets of Lambdas, called the Causaloid ${\bf \Lambda}$. In the follow-up work \cite{sakharwade2023hierarchy} we will study Meta Compression in further detail, providing a class of identities for Compositional Lambdas that will provide a way to classify physical theories.  

\subsubsection{The Causaloid Product}

Let us discuss products. Within Quantum theory there are three basic ways of putting two operators together in quantum theory.  Two are well known.  We have $\hat A \hat B$ (sequential, temporal and causally adjacent) when, for example, a qubit passes through a box associated with $\hat{A}$ then immediately afterwards, passes through a box associated with $\hat{A}$ (sometimes this product is denoted $\hat{A}\circ\hat{B}$).  We have $\hat A\otimes \hat B$ (spatially separate) when, for example, two qubits each pass through space-like separated boxes associated with $\hat{A}$ and $\hat{B}$.  We can also define the less familiar product $\hat A ? \hat B$ (temporal but not causally adjacent) when, for example, a qubit passes through a box associated with $\hat{A}$, then a box associated with some other operator (call it $\hat{C}$) and then passes through a box associated with $\hat{B}$.   In this case we can define $\hat A ? \hat B$ through the relationship $[\hat A ? \hat B] \hat{C} = \hat{A}\hat{C}\hat{B}$ for all $\hat{C}$.  One of the goals of the Causaloid framework, with an interest in studying indefinite causality, is to be able to treat space and time on an equal footing and thus to treat these different kinds of products on an equal footing. To this end the {\it Causaloid product} is defined. While this is in a general framework, the Causaloid product can be shown to unify the different types of
products within quantum theory.  

Given ${\bf r}_{\alpha_1}$ measurement vector in $R_{\mathcal{O}_1}$ (or simply $R_{1}$) and ${\bf r}_{\alpha_2}$ measurement vector in $R_{\mathcal{O}_2}$ (or simply $R_{2}$) such that $R_{\mathcal{O}_1}\cap R_{\mathcal{O}_2} =\phi$. The Causaloid product $\otimes^{\Lambda}$ is defined as:
\begin{equation}\label{causaloidprodO}
{\bf r}_{\alpha_1\alpha_2}({\cal O}_1\cup{\cal O}_2) = {\bf r}_{\alpha_1}({\cal O}_1) \otimes ^{\Lambda} {\bf
r}_{\alpha_2}({\cal O}_2)
\end{equation}
or simply
\begin{equation}\label{causaloidprod}
{\bf r}_{\alpha_1\alpha_2}= {\bf r}_{\alpha_1}\otimes ^{\Lambda} {\bf r}_{\alpha_2}
\end{equation}
where the labels $\alpha$ implicitly specify the regions. 
Diagrammatically we represent the Causaloid product with the symbol $\otimes^{\Lambda}$:
\begin{align}\label{causaloidproddiagram}
\begin{DiagramV}[0.8]{0}{0}
\begin{move}{0,0}
\draw (0,4) -- (6,4); 
\draw (0,-4) -- (6,-4); 
\draw (6,-7) -- (6,-2);
\fill[blue!80!red!20!] (6,0) ellipse (2 and 5);
\draw (6,0) ellipse (2 and 5) node {${\bf r}$};
\draw (2.5,4+1) node {\footnotesize $\alpha_{1}$};
\draw (2.5,-4+1) node {\footnotesize $\alpha_{2}$};
\fill[black] (6,-6) circle (0.25);
\end{move}
\end{DiagramV}
=
\begin{DiagramV}[0.8]{0}{0}
\begin{move}{0,0}
\draw (0,4) -- (6,4); 
\draw (0,-4) -- (6,-4); 
\draw  (6,4) to[out=0,in=90](10,0);
\draw  (6,-4) to[out=0,in=270](10,0);
\fill[blue!80!red!20!] (6,4) circle (2);
\draw (6,4) circle (2) node {${\bf r}$};
\fill[blue!80!red!20!] (6,-4) circle (2);
\draw (6,-4) circle (2) node {${\bf r}$};
\draw (2.5,4+1) node {\footnotesize $\alpha_{1}$};
\draw (2.5,-4+1) node {\footnotesize $\alpha_{2}$};
\end{move}
\begin{scope}[shift={(10,0)}]
\draw  (0,0) to[out=0,in=90](4,-6);
\draw (4,-6) -- (4,-7);
\fill[black] (4,-6) circle (0.25);
\fill[white] (0,0) circle (0.5);
\draw (0,0) node {\small $\mathbf{\otimes}$};
\draw (0.85,0.6) node {\small $\mathbf{^\Lambda}$};
\end{scope}
\end{DiagramV}
\end{align}

The way to expand the Causaloid product is given by Lambda matrices which in turn are given by the Causaloid. The component form for Equations \eqref{causaloidprodO}-\eqref{causaloidproddiagram} using Equation \eqref{rsecondlevelcomponent} is given by:
\begin{equation}\label{genproduct}
{\bf r}_{\alpha_1\alpha_2}|_{k_1k_2} = \sum_{l_1l_2\in\Omega_1\times\Omega_2} ({\bf r}_{\alpha_1}|_{l_1}) ({\bf
r}_{\alpha_2}|_{l_2}) \Lambda_{l_1l_2}^{k_1k_2}
\end{equation}
Diagrammatically we can see that the Causaloid product may be expanded as follows:
\begin{align}
\begin{DiagramV}[0.8]{0}{0}
\begin{move}{0,0}
\draw (0,4) -- (6,4); 
\draw (0,-4) -- (6,-4); 
\draw  (6,4) to[out=0,in=90](10,0);
\draw  (6,-4) to[out=0,in=270](10,0);
\fill[blue!80!red!20!] (6,4) circle (2);
\draw (6,4) circle (2) node {${\bf r}$};
\fill[blue!80!red!20!] (6,-4) circle (2);
\draw (6,-4) circle (2) node {${\bf r}$};
\draw (2.5,4+1) node {\footnotesize $\alpha_{1}$};
\draw (2.5,-4+1) node {\footnotesize $\alpha_{2}$};
\end{move}
\begin{scope}[shift={(10,0)}]
\draw  (0,0) to[out=0,in=90](4,-6);
\draw (4,-6) -- (4,-7);
\fill[black] (4,-6) circle (0.25);
\fill[white] (0,0) circle (0.5);
\draw (0,0) node {\small $\mathbf{\otimes}$};
\draw (0.85,0.6) node {\small $\mathbf{^\Lambda}$};
\end{scope}
\end{DiagramV}
=
    \begin{DiagramV}[0.8]{0}{0}
    \begin{move}{0,0}
    \draw (0,4) to[out=0,in=180] (4,1); 
    \draw (0,-4) to[out=0,in=180] (4,-1); 
    \draw (6,0) -- (6,-7); 
     \draw (-8,-4) -- (0,-4); 
    \fill[black] (6,0) circle (0.25);
    \fill[blue!80!red!20!] (6,0) ellipse (2 and 3);
    \draw (6,0) ellipse (2 and 3) node {$p$};
    \fill[blue!80!red!20!] (6,0) ellipse (2 and 3);
    \draw (6,0) ellipse (2 and 3) node {${\bf r}$};
    \fill[black] (6,-6) circle (0.25);
    \draw (-8,4) -- (0,4); 
    \fill[blue!80!red!20!] (-4,-2-4) rectangle (0,2+4);
    \draw (-4,-2-4) rectangle (0,2+4);
    \draw (-2,0) node {$\Lambda$};
    \draw (-6,1+4) node {\footnotesize $l_{1}$};
    \draw (-6,1-4) node {\footnotesize $l_{2}$};
    \draw (2.5,4) node {\footnotesize $k_{1}$};
    \draw (2.5,-4) node {\footnotesize $k_{2}$};
    \end{move}
\begin{scope}[shift={(-8,0)}]
    \draw (0,4) -- (4,4); 
    \draw (0,-4) -- (4,-4); 
    \draw (-8,4) -- (0,4); 
    \fill[blue!80!red!20!] (-4,-2+4) rectangle (0,2+4);
    \draw (-4,-2+4) rectangle (0,2+4);
    \draw (-2,0+4) node {$\Lambda$};
    \draw (-6,1+4) node {\footnotesize $\alpha_{1}$};
    \draw (-8,-4) -- (0,-4); 
    \fill[blue!80!red!20!] (-4,-2-4) rectangle (0,2-4);
    \draw (-4,-2-4) rectangle (0,2-4);
    \draw (-2,0-4) node {$\Lambda$};
    \draw (-6,1-4) node {\footnotesize $\alpha_{2}$};
    \end{scope}
    \end{DiagramV}
\end{align}
This can be massaged into a more illuminating form to identify the kinds of different products the unified Causaloid product can reduce to: 
\begin{align}\label{rtwoterms}
{\bf r}_{\alpha_1\alpha_2}|_{k_1k_2} &= \sum_{l_1l_2\in\Omega_1\times\Omega_2} ({\bf r}_{\alpha_1}|_{l_1}) ({\bf
r}_{\alpha_2}|_{l_2}) \Lambda_{l_1l_2}^{k_1k_2}\\
&= \sum_{l_1l_2\in\Omega_{1,2}}({\bf r}_{\alpha_1}|_{l_1}) ({\bf r}_{\alpha_2}|_{l_2}) \delta_{l_1l_2}^{k_1k_2} +
\sum_{l_1l_2\in\Omega_1\times\Omega_2-\Omega_{1,2}} ({\bf r}_{\alpha_1}|_{l_1}) ({\bf r}_{\alpha_2}|_{l_2}) \Lambda_{l_1l_2}^{k_1k_2}\\
&= ({\bf r}_{\alpha_1}|_{k_1}) ({\bf r}_{\alpha_2}|_{k_2}) +
\sum_{l_1l_2\in\Omega_1\times\Omega_2-\Omega_{1,2}} ({\bf r}_{\alpha_1}|_{l_1}) ({\bf r}_{\alpha_2}|_{l_2}) \Lambda_{l_1l_2}^{k_1k_2}
\end{align}
where we first expand the sum into two terms over different sets since $\Omega_{1,2} +(\Omega_1 \times\Omega_2 -\Omega_{1,2}) = \Omega_1 \times\Omega_2$, and in the second step we use the relation
\begin{align}
    \Lambda_{l_1l_2}^{k_1k_2}=\delta_{l_1l_2}^{k_1k_2}\quad \text{for}\quad l_1,l_2 \in \Omega_{1,2}
\end{align}

There are two special cases for Equation \eqref{rtwoterms} given Equation \eqref{centralresult}.
\begin{enumerate}
\item {\it No Compositional Compression} $\Omega_{1,2}=\Omega_1\times\Omega_2$ such that
$|\Omega_{1,2}|=|\Omega_1||\Omega_2|$.
\item {\it Non-trivial Compositional Compression} $\Omega_{1,2}\subset\Omega_1\times\Omega_2$ such that $|\Omega_{1,2}|< |\Omega_1||\Omega_2|$.
\end{enumerate}

It follows that
\begin{equation}\label{ifomegamultiply}
{\rm if}~~~\Omega_{1,2}=\Omega_1\times\Omega_2~~~{\rm then}~~~ {\bf r}_{\alpha_1\alpha_2} = {\bf
r}_{\alpha_1}\otimes{\bf r}_{\alpha_2}
\end{equation}
where $\otimes$ stands for the ordinary tensor product. Hence the ordinary tensor product is a special case of the Causaloid product. 

We will see in the follow-up work \cite{sakharwade2023duotensors} through the duotensor formalism, that in (finite-dimensional) quantum theory, typically $\Omega_{1,2}=\Omega_1\times \Omega_2$ corresponds to the products $\hat A\otimes \hat B$ as well as $\hat A ? \hat B$. Since the total number of real parameters after taking the product is equal to the product of the number from each operator we have $|\Omega_{1,2}|=|\Omega_1||\Omega_2|$.  

Non-trivial Compositional compression occurs ($|\Omega_{1,2}|<|\Omega_1||\Omega_2|$) when two regions are causally adjacent, such that what happens in one region depends, at least partially, on what is done in the other region in a way that cannot be tampered by other regions in $R$. In quantum theory when we take the product $\hat A\hat B$, the total number of real parameters in the product is equal to the number in $\hat A$ (or $\hat B$), giving the special case $|\Omega_{1,2}|=|\Omega_1|=|\Omega_2|$. 

Therefore, the Causaloid product is expected to unify all the different types of products a physical theory will have to offer, which can be calculated once the Causaloid is specified. This concludes the discussion of the three levels of compression.  It is worth noting, in this respect, that the product of two quantum combs \cite{chiribella2009theoretical} can be understood in this language (indeed, Chribella et al.\ note in Ref \cite{chiribella2009theoretical} \lq\lq a striking analogy between quantum combs and a quantum realisation of the Causaloid\rq\rq).

\subsection{\label{synopsis}Synopsis: Three Levels of Physical Compression}

We presented new diagrammatics for the Causaloid framework's three levels of physical compression. Here we provide a short and handy synopsis for quick reference. 

We considered a large region $R$ which is predictively well-defined. We focus on disjoint regions $R_1, R_2,...$ within $R$. We expect the framework to help us predict probabilities. Using the generalised Born rule given linearly in terms of states and measurements gives probabilities. 
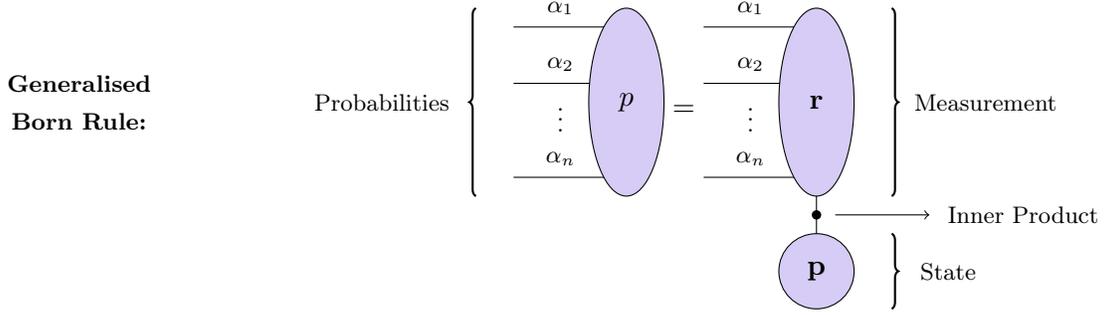
\begin{align}
\begin{DiagramV}{3}{0}
\draw (0,+1) node {\footnotesize {\bf Generalised}};
\draw (0,-1) node {\footnotesize {\bf Born Rule:}};
\end{DiagramV}
\begin{DiagramV}{0}{0}
\begin{move}{0,0}
\draw (0,4) -- (6,4); 
\draw (0,1) -- (6,1); 
\draw (0,-4) -- (6,-4); 
\fill[blue!80!red!20!] (6,0) ellipse (2 and 5);
\draw (6,0) ellipse (2 and 5) node {$p$};
\draw (2.5,4+1) node {\footnotesize $\alpha_{1}$};
\draw (2.5,1+1) node {\footnotesize $\alpha_{2}$};
\draw (2.5,-1.5+1) node {\footnotesize $\vdots$};
\draw (2.5,-4+1) node {\footnotesize $\alpha_{n}$};
\draw (-7,0) node {\footnotesize Probabilities};
\draw [thick,decorate,
    decoration = {brace}] (-2,-5) --  (-2,5);
\end{move}
\end{DiagramV}
=
\begin{DiagramV}{0}{0}
\begin{move}{0,0}
\draw (0,4) -- (6,4); 
\draw (0,1) -- (6,1); 
\draw (0,-4) -- (6,-4); 
\draw (6,-9) -- (6,-2);
\fill[blue!80!red!20!] (6,0) ellipse (2 and 5);
\draw (6,0) ellipse (2 and 5) node {${\bf r}$};
\draw (2.5,4+1) node {\footnotesize $\alpha_{1}$};
\draw (2.5,1+1) node {\footnotesize $\alpha_{2}$};
\draw (2.5,-1.5+1) node {\footnotesize $\vdots$};
\draw (2.5,-4+1) node {\footnotesize $\alpha_{n}$};
\fill[black] (6,-6) circle (0.25);
\fill[blue!80!red!20!] (6,-9) circle (2);
\draw (6,-9) circle (2) node {${\bf p}$};
\draw (15,0) node {\footnotesize Measurement};
\draw [thick,decorate,
    decoration = {brace}] (10,5) --  (10,-5);
\draw (13,-9) node {\footnotesize State};
\draw [thick,
    decoration={brace},decorate] (10,-7) --  (10,-11);
\draw[->] (7,-6) -- (12,-6);
\draw (17,-6) node {\footnotesize Inner Product};
\end{move}
\end{DiagramV}
\nonumber
\end{align}
The ${\bf r}$ measurement vector can be compressed. Tomographic Compression concerns a single region and Compositional Compression concerns multiple disjoint regions.   
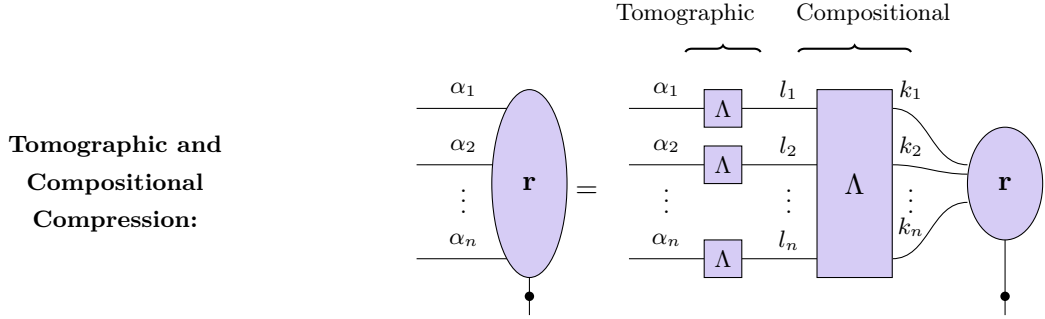
\begin{align}
\begin{DiagramV}{4}{0}
\draw (0,+2) node {\footnotesize {\bf Tomographic and}};
\draw (0,+0) node {\footnotesize {\bf Compositional}};
\draw (0,-2) node {\footnotesize {\bf Compression:}};
\end{DiagramV}
\begin{DiagramV}{0}{0}
\begin{move}{0,0}
\draw (0,4) -- (6,4); 
\draw (0,1) -- (6,1); 
\draw (0,-4) -- (6,-4); 
\draw (6,-7) -- (6,-2);
\fill[blue!80!red!20!] (6,0) ellipse (2 and 5);
\draw (6,0) ellipse (2 and 5) node {${\bf r}$};
\draw (2.5,4+1) node {\footnotesize $\alpha_{1}$};
\draw (2.5,1+1) node {\footnotesize $\alpha_{2}$};
\draw (2.5,-1.5+1) node {\footnotesize $\vdots$};
\draw (2.5,-4+1) node {\footnotesize $\alpha_{n}$};
\fill[black] (6,-6) circle (0.25);
\end{move}
\end{DiagramV}
=
\begin{DiagramV}{3.5}{0}
\begin{scope}[shift={(-8,0)}]
\draw (-6,4) -- (6,4); 
\draw (-6,1) -- (6,1); 
\draw (-6,-4) -- (6,-4); 
\draw (2.5,4+1) node {\footnotesize $l_{1}$};
\draw (2.5,1+1) node {\footnotesize $l_{2}$};
\draw (2.5,-1.5+1) node {\footnotesize $\vdots$};
\draw (2.5,-4+1) node {\footnotesize $l_{n}$};
\draw (-4,4+1) node {\footnotesize $\alpha_{1}$};
\draw (-4,1+1) node {\footnotesize $\alpha_{2}$};
\draw (-4,-1.5+1) node {\footnotesize $\vdots$};
\draw (-4,-4+1) node {\footnotesize $\alpha_{n}$};
%
\fill[blue!80!red!20!] (-2,-1+4) rectangle (0,1+4);
\draw (-2,-1+4) rectangle (0,1+4);
\draw (-1,+4) node {\footnotesize $\Lambda$};
\fill[blue!80!red!20!] (-2,-1+1) rectangle (0,1+1);
\draw (-2,-1+1) rectangle (0,1+1);
\draw (-1,+1) node {\footnotesize $\Lambda$};
\fill[blue!80!red!20!] (-2,-1-4) rectangle (0,1-4);
\draw (-2,-1-4) rectangle (0,1-4);
\draw (-1,-4) node {\footnotesize $\Lambda$};
\draw (-3,+9) node {\footnotesize Tomographic};
\draw [thick,decorate,
    decoration = {brace}] (-3,7) --  (1,7);
\end{scope}
\begin{move}{0,0}
\draw (0,4) to[out=0,in=180] (4,1); 
\draw (0,1) to[out=0,in=180] (4,0.5); 
\draw (0,-4) to[out=0,in=180] (4,-1); 
\draw (6,0) -- (6,-7); 
\fill[black] (6,0) circle (0.25);
\fill[blue!80!red!20!] (6,0) ellipse (2 and 3);
\draw (6,0) ellipse (2 and 3) node {${\bf r}$};
\fill[black] (6,-6) circle (0.25);
%
\fill[blue!80!red!20!] (-4,-1-4) rectangle (0,1+4);
\draw (-4,-1-4) rectangle (0,1+4);
\draw (-2,0) node {$\Lambda$};
\draw (-1,+9) node {\footnotesize Compositional};
\draw [thick,decorate,
    decoration = {brace}] (-5,7) --  (1,7);
\draw (1,5) node {\footnotesize $k_{1}$};
\draw (1,2) node {\footnotesize $k_{2}$};
\draw (1,-0.5) node {\footnotesize $\vdots$};
\draw (1,-3+0.7) node {\footnotesize $k_{n}$};
\end{move}
\end{DiagramV}\nonumber
\end{align}
The label change between these levels of compression is given by:
\begin{align}
\quad
   &\alpha_1,\alpha_2,\dots \alpha_n \in \Gamma_1\times\Gamma_2\times\dots\Gamma_n \nonumber\\
   \longrightarrow\quad& l_1,l_2,\dots l_n \in \Omega_1\times\Omega_2\times\dots\Omega_n \nonumber\\
   \longrightarrow\quad& k_1,k_2,\dots k_n \in \Omega_{1,2,\dots n} \nonumber
\end{align}

Given the first two levels of Compression, we get the set of Lambdas for any region in $R$ which helps us specify \emph{the Causaloid} ${\bf \Lambda}$, which may be considered as ``a specification of a physical theory itself\rq\rq.

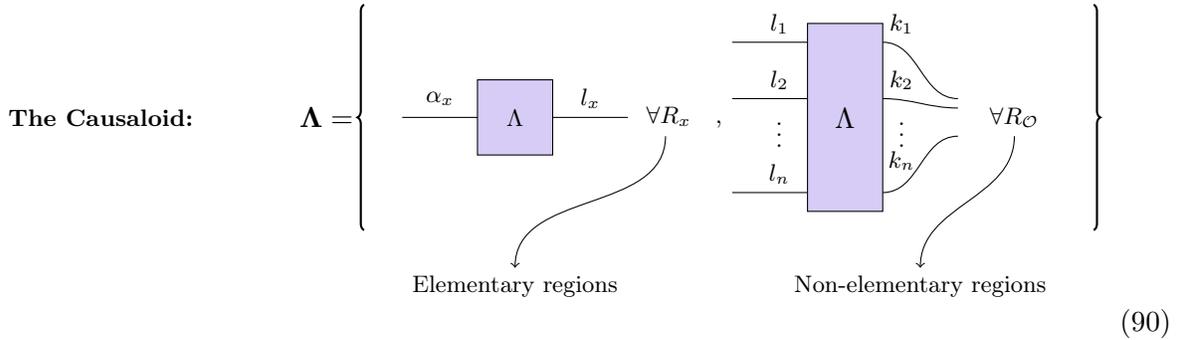
\begin{align}
    \begin{DiagramV}{4}{0}
    \draw (0,+0) node {\footnotesize {\bf The Causaloid:}};
    \draw (12,0) node {${\bf \Lambda=}$};  
    \draw [thick,decorate,
        decoration = {brace}] (14,-6) --  (14,6);
    \end{DiagramV}
    \begin{DiagramV}{0}{0}
    \draw (-6,0) -- (6,0); 
    \draw (4,1) node {\footnotesize $l_{x}$};
    \draw (-4,1) node {\footnotesize $\alpha_{x}$};
    \fill[blue!80!red!20!] (-2,-2) rectangle (2,2);
    \draw (-2,-2) rectangle (2,2);
    \draw (-0,+0) node {\footnotesize $\Lambda$};
    %
    \draw[->] (8,-1) to[out=270,in=90] (0,-8); 
    \draw (0,-9) node {\footnotesize Elementary regions};
    \draw (9,0) node {\footnotesize $\forall R_x\quad$,};
    \end{DiagramV}
    \begin{DiagramV}{0}{0}
    \begin{scope}[shift={(-8,0)}]
    \draw (-0,4) -- (6,4); 
    \draw (-0,1) -- (6,1); 
    \draw (-0,-4) -- (6,-4); 
    \draw (2.5,4+1) node {\footnotesize $l_{1}$};
    \draw (2.5,1+1) node {\footnotesize $l_{2}$};
    \draw (2.5,-1.5+1) node {\footnotesize $\vdots$};
    \draw (2.5,-4+1) node {\footnotesize $l_{n}$};
    \end{scope}
    \begin{move}{0,0}
    \draw (0,4) to[out=0,in=180] (4,1); 
    \draw (0,1) to[out=0,in=180] (4,0.5); 
    \draw (0,-4) to[out=0,in=180] (4,-1); 
    \fill[blue!80!red!20!] (-4,-1-4) rectangle (0,1+4);
    \draw (-4,-1-4) rectangle (0,1+4);
    \draw (-2,0) node {$\Lambda$};
    %
    \draw (7,0) node {\footnotesize $\forall R_{\mathcal{O}}$};
    \draw[->] (7,-1) to[out=270,in=90] (2,-8); 
    \draw (2,-9) node {\footnotesize Non-elementary regions};
    \draw (1,5) node {\footnotesize $k_{1}$};
    \draw (1,2) node {\footnotesize $k_{2}$};
    \draw (1,-0.5) node {\footnotesize $\vdots$};
    \draw (1,-3+0.7) node {\footnotesize $k_{n}$};
    \end{move}
    \end{DiagramV}
    \begin{DiagramV}{4}{0}
    \draw [thick,decorate,
        decoration = {brace}] (2,6) --  (2,-6);
    \end{DiagramV}
\end{align}

The third level or Meta Compression provides us with \emph{rules} to be able to calculate any Lambda Matrix given a reduced set which gives a shorter specification of the Causaloid. The Causaloid in turn helps us define the Causaloid Product $\otimes^\Lambda$, which unifies different kinds of spatiotemporal products within a theory and gives us any general ${\bf r}$-vectors.
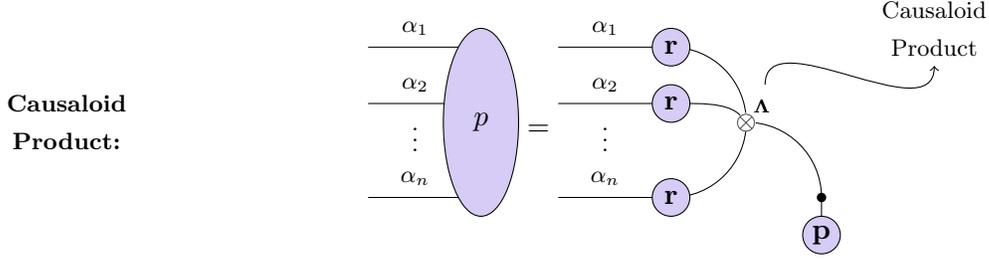
\begin{align}
\begin{DiagramV}{4}{0}
\draw (0,+1) node {\footnotesize {\bf Causaloid}};
\draw (0,-1) node {\footnotesize {\bf Product:}};
\end{DiagramV}
\begin{DiagramV}{0}{0}
\begin{move}{0,0}
\draw (0,4) -- (6,4); 
\draw (0,1) -- (6,1); 
\draw (0,-4) -- (6,-4); 
\fill[blue!80!red!20!] (6,0) ellipse (2 and 5);
\draw (6,0) ellipse (2 and 5) node {$p$};
\draw (2.5,4+1) node {\footnotesize $\alpha_{1}$};
\draw (2.5,1+1) node {\footnotesize $\alpha_{2}$};
\draw (2.5,-1.5+1) node {\footnotesize $\vdots$};
\draw (2.5,-4+1) node {\footnotesize $\alpha_{n}$};
%
\end{move}
\end{DiagramV}
=
\begin{DiagramV}{7}{0}
\begin{move}{0,0}
\draw (0,4) -- (6,4); 
\draw (0,1) -- (6,1); 
\draw (0,-4) -- (6,-4); 
\draw  (6,4) to[out=0,in=90](10,0);
\draw  (6,1) to[out=0,in=90](10,-0.5);
\draw  (6,-4) to[out=0,in=270](10,0);
\fill[blue!80!red!20!] (6,4) circle (1);
\draw (6,4) circle (1) node {${\bf r}$};
\fill[blue!80!red!20!] (6,1) circle (1);
\draw (6,1) circle (1) node {${\bf r}$};
\fill[blue!80!red!20!] (6,-4) circle (1);
\draw (6,-4) circle (1) node {${\bf r}$};
\draw (2.5,4+1) node {\footnotesize $\alpha_{1}$};
\draw (2.5,1+1) node {\footnotesize $\alpha_{2}$};
\draw (2.5,-1.5+1) node {\footnotesize $\vdots$};
\draw (2.5,-4+1) node {\footnotesize $\alpha_{n}$};
\end{move}
\begin{scope}[shift={(10,0)}]
\draw  (0,0) to[out=0,in=90](4,-4);
\draw (4,-4) -- (4,-6);
\fill[black] (4,-4) circle (0.25);
\fill[blue!80!red!20!] (4,-6) circle (1);
\draw (4,-6) circle (1) node {${\bf p}$};
\fill[white] (0,0) circle (0.5);
\draw (0,0) node {\small $\mathbf{\otimes}$};
\draw (0.85,0.6) node {\small $\mathbf{^\Lambda}$};
\draw (10,6) node {\footnotesize Causaloid};
\draw (10,4) node {\footnotesize Product};
\draw[->]  (1,2) to[out=90,in=270](10,3);
\end{scope}
\end{DiagramV}
\nonumber
\end{align}
This brings us full circle. The framework prescribes a way to organise recorded data using the structure of physical theories, by distilling down to a compressed version and studying correlations. 

\section{Prediction Heralding} \label{sec:calculatingprobabilities}

Our objective, as stated in Section \ref{subsec:welldefinedprobabilities} is to be able to do {\it prediction heralding} and, when probabilities are heralded to be well defined, actually calculate them.  The general form of probability we want to be able to calculate is
\begin{equation}  \label{probYonegivenYtwo}
\text{prob}(Y_1^{\alpha_1}|Y_2^{\alpha_2}, F_1^{\alpha_1}, F_2^{\alpha_2}) 
\end{equation}
First, we wish to test whether this is well defined.  So let us consider, instead, the probability
\begin{equation}  \label{probYonegivenYtwoplus}
\text{prob}(Y_1^{\alpha_1}|Y_2^{\alpha_2}, Y_{R-R_1-R_2}, F_1^{\alpha_1}, F_2^{\alpha_2}, F_{R-R_1-R_2}) 
=
\frac{({\bf r}^{\alpha_1} \otimes {\bf r}^{\alpha_2})\cdot {\bf p}_{R-R_1-R_2}}{\sum\limits_{\alpha'_1 \sim F_1^{\alpha_1} } ({\bf r}^{\alpha'_1} \otimes {\bf r}^{\alpha_2})\cdot {\bf p}_{R-R_1-R_2} }
\end{equation}
where $\alpha'_1 \sim F_1^{\alpha_1}$ in the sum means we sum over all $\alpha'_1$ that are consistent with the procedure, $F_1^{\alpha_1}$ in $R_1$ (this means we are summing over all outcomes for this region).  The probability in \eqref{probYonegivenYtwo} is well defined if and only if the probability in \eqref{probYonegivenYtwoplus}  is independent of what happens in $R-R_1-R_2$.  This, in turn, is true if and only if
\begin{equation}\label{heraldingcondition}
   {\bf r}^{\alpha_1} \otimes {\bf r}^{\alpha_2} = p  \sum\limits_{\alpha'_1 \sim F_1^{\alpha_1} } {\bf r}^{\alpha'_1} \otimes {\bf r}^{\alpha_2}
\end{equation}
for some $p$
(i.e.\ if these two vectors are parallel).  This follows since this is the necessary and sufficient condition for the probability being calculated in \eqref{probYonegivenYtwoplus} not to depend on ${\bf p}_{R-R_1-R_2}$ (the state associated with $R-R_1-R_2$).  This is the prediction heralding condition.  If this condition is satisfied then we can say that the probability in \eqref{probYonegivenYtwo} is equal to the constant of proportionality, $p$ in \eqref{heraldingcondition}.

It is worth recalling, as discussed in Section \ref{subsec:welldefinedprobabilities} that the possibility of indefinite causal structure forces us to adopt a heralding approach since we cannot rely on background causal structure to tell us which probabilities are well-defined.

\section{\label{sec:causaloiddiscussion}Outlook and conclusions}

In this paper, we have revisited the Causaloid framework and provided a new diagrammatic representation for the three levels of physical compression.  In part, our objective is that this diagrammatic review will make the Causaloid framework more accessible to those interested, who are engaged in studying indefinite causal structures.  

In future follow-up works, we will utilise the diagrammatics presented here, to study Meta-Compression further giving us a Hierarchy \cite{sakharwade2023hierarchy}; and to diagrammatically apply the Causaloid framework to the Duotensor framework (within which we can formulate Classical Probability Theory and Quantum Theory), placing it into the second rung of a hierarchy of theories \cite{sakharwade2023duotensors} wherein it is only necessary to specify Lambda matrices for pairs of elementary regions to specify the Casualoid.   

\section{Acknowledgements}

This work was supported through grants from the John Templeton Foundation funding for phases 1 and 2 of the \lq\lq The Quantum Information Structure of Spacetime (QISS)\rq\rq\: project as well as a Discovery Grant from the Natural Sciences and Engineering Research Council of Canada.  
Research at Perimeter Institute is supported in part by the Government of Canada through the Department of Innovation, Science and Economic Development Canada and by the Province of Ontario through the Ministry of Colleges and Universities.
N.S. acknowledges partial support by the Foundation for Polish Science -- IRAP project, ICTQT, contract no. MAB/2018/5, as well as, partial support by the PNRR MUR Project No. CN 00000013-ICSC (Spoke 10 - Quantum Computing).

\bibliographystyle{ieeetr}
\bibliography{Draft-1.bib}
 
\end{document}